\documentclass[a4paper,single column,12pt]{article}
\usepackage[colorlinks=true,linktocpage=true,linkcolor=blue,citecolor=blue]{hyperref}
\usepackage{graphicx}
\usepackage{dcolumn}
\usepackage{bm}
\usepackage{subfig}
\usepackage{float}
\usepackage{tabularx}
\usepackage{hhline}
\usepackage{color}
\usepackage{bigints}
\usepackage{epsfig,amsmath,amsfonts,amsthm,amssymb,amstext,amsbsy}
\usepackage{mathtools,slashed}
\usepackage[most]{tcolorbox}
\usepackage[toc,page]{appendix}
\usepackage{chngpage}
\usepackage[english]{babel}

\newcommand{\old}[1]{}

\newcommand\ems{(e^- e^+ \rightarrow \mu^- \mu^+)}
\newcommand\emt{(e^- \mu^- \rightarrow e^- \mu^-)}
\newcommand\bhabha{(e^- e^+ \rightarrow e^- e^+)}
\newcommand\moller{(e^- e^- \rightarrow e^- e^-)}
\usepackage{caption}
\begin{document}
	\newcommand{\g}{\gamma}
    \newcommand{\s}{\slashed}
	\newcommand{\Tr}{\mathop{\rm Tr}\nolimits}
	\newcommand{\para}{_\parallel}
	\newcommand{\pr}{_\perp}
	\newcommand{\fs}{\rlap/}
	\def\twidle{\widetilde}
	\def\f{\frac}
	\def\omit#1{_{\!\rlap{$\scriptscriptstyle \backslash$}
			{\scriptscriptstyle #1}}}
	\def\vec#1{\mathchoice 
		{\mbox{\boldmath $#1$}}
		{\mbox{\boldmath $#1$}}
		{\mbox{\boldmath $\scriptstyle #1$}}
		{\mbox{\boldmath $\scriptscriptstyle #1$}}
	}
	\def\eqn#1{Eq.\ (\ref{#1})}

\begin{flushright}
{\normalsize
}
\end{flushright}
\vskip 0.1in
\begin{center}
{\large {\bf
Lowest-order electron-electron and electron-muon scattering 
in a strong magnetic field}}
\end{center}
\vskip 0.1in
\begin{center}
Abhishek Tiwari$^\dag$\footnote{abhi7phy@gmail.com} and Binoy Krishna 
Patra$^\dag$ \footnote{binoyfph@iitr.ac.in}
\vskip 0.02in
{\small {\it $^\dag$ Department of Physics, Indian Institute of
Technology Roorkee, Roorkee 247 667, India\\
} }
\end{center}
\vskip 0.01in
\addtolength{\baselineskip}{0.4\baselineskip} 
\section* {Abstract}
In this work we have investigated how the much studied 
scattering processes in vacuum at the lowest-order,
{\em viz.} electron-muon ($e$-$\mu$) scattering in both $s$- and 
$t$-channel, Bhabha scattering, and M\o ller scattering, 
have been modified in the presence of a strong magnetic field 
($|eB|>>m^2$, $m$ is the mass of electron or muon). 
For that purpose, we have first calculated the square of the matrix 
element by summing over the spin states, using the spinors 
in the presence of a strong magnetic field and then obtain the crosssection
by integrating over the available phase space for the final
states and averaging over the initial states. The first noticeable
observation in the spin-summed of the matrix element
squared is that the interference term between $s$- and $t$-channel 
and $t$- and $u$-channel in Bhabha and M\o ller scattering, 
respectively 
are missing in the presence of strong magnetic field.
We have found that in the presence of strong magnetic field, 
the crosssection of $e^- e^+ \rightarrow \mu^- \mu^+$ annihilation 
process
in the lowest order decreases inversely proportional to the fourth 
power 
of the center-of-mass energy ($\sqrt{s}$), compared to the 
inversely proportional to the square of center-of-mass energy 
in vacuum alone. Like in vacuum, 
the crosssection for $e$-$\mu$ scattering in ${\rm t}$ channel, 
{\em i.e.} for $e^- \mu^- \rightarrow e^- \mu^-$ process, even in a strong 
magnetic field too diverges but the finite part
decreases with $\sqrt{s}$ much faster than in vacuum alone.
 Similarly the crosssections 
for both Bhabha and M\o ller scattering at the lowest-order 
diverges in the infrared limit.
However, the finite term decreases with $\sqrt{s}$
much faster than the vacuum alone. In addition there is finite negative 
contribution, which is independent of $\sqrt{s}$ and decreases with the 
magnetic field.

\noindent PACS:~~ 12.39.-x,11.10.St,12.38.Mh,12.39.Pn
12.75.N, 12.38.G \\
\vspace{1mm}
\noindent{\bf Keywords}: Dirac spinor, electron-electron scattering, 
electron-muon scattering, strong magnetic field, crosssection, 
Mandelstam and magnetic Mandelstam variables\\

\section{Introduction}
The relativistic heavy-ion collider (RHIC) at the Brookhaven National 
Laboratory, USA, with the center-of-mass energy, $\sqrt{s}$= 200 GeV per
nucleon in Au - Au collisions and the large hadron
collider (LHC) at the European Organization for Nuclear Research,
Geneva, with $\sqrt{s}$ = 2.76 TeV per nucleon in Pb-Pb collisions may have
produced intensely strong magnetic field
at very early stages of collisions, when the event is off-central
\cite{Shovkovy:LNP871_2013,Elia:LNP871_2013,
Fukushima:LNP871_2013,Mulller:PRD89_2014,
Miransky:PR576_2015}. 
Depending on the centralities, the strength of the magnetic field may
reach between $m_{\pi}^2$ ($\simeq 10^{18}$ Gauss) at
RHIC \cite{Kharzeev:NPA803_2008} to 15 $m_{\pi}^2$ at
LHC \cite{Skokov:IJMPA24_2009}.
At extreme cases it may reach values of 50 $m_{\pi}^2$ at LHC.
A very strong magnetic field ($\sim 10^{23}$~Gauss) may have existed 
in the early universe during the electroweak
phase transition due to the gradients in Higgs field
\cite{vachaspati:PLB265'1991} or at the core of magnetars~\cite{magnetars}.

Thus we are motivated in this work to study the different processes 
in electron-muon and electron-electron scattering in the lowest order in the 
presence of strong magnetic 
field. However, the above processes in the lowest as well as higher 
order are well studied theoretically in 
vacuum~\cite{bonciani_ferroglia,ilyichev_zykunov,halzen_martin} with 
the solutions of Dirac equation in vacuum, {\em i.e.} with the free Dirac 
spinors for positive and negative energy and their 
corresponding completeness relation.
 However, the scenario in the
 presence of strong magnetic field is different because the form of
the Dirac
 spinors are going to change, where, apart from the momentum dependence, 
 the spinors also depend on the spatial coordinates due to the
 gauge used to solve the Dirac equation in magnetic field. As a result,
the completeness relations are also going to change.
 For the sake of simplicity we assume the magnetic field to be uniform 
and stationary. Moreover the strength of the magnetic field is
strong enough so that 
only the lowest Landau level is sufficient for the calculation
of the matrix element and the crosssection.

This paper is divided into following sections. We first 
nomenclature the
four-momentum, Mandelstam variables, magnetic Mandelstam variables,
suitable for the description in a strong magnetic field and then 
revisit the Dirac equation in a strong and homogeneous magnetic 
field in
section 2.1 and 2.2, respectively. Using those notations, 
we calculate the spin-summed matrix element squared for the 
 electron-muon scattering $s$-channel (annihilation process) and $t$- 
 channel, Bhabha scattering and M\o ller scattering in sections 
 2.3-2.5, 
respectively. In section 3, we first revisit for the formula 
for calculating the crosssection by constructing the Lorentz 
invariant phase space, flux factor, energy-momentum conserving 
Dirac-Delta
function etc. in the presence of strong magnetic field and then 
using the
the matrix elements for the above processes from the above sections 2.3-2.5,
the corresponding crosssections have been evaluated in sections 3.1-3.4, 
respectively. Finally we  conclude our results and discussion 
in section 4.

\section{Electron-Electron and Electron-Muon Scattering in Strong Magnetic 
field}
Our aim in this section is to calculate the square of matrix
element for the electron-muon scattering in both $s$- and 
$t$-channel,
Bhabha scattering, and M\o ller scattering
in a strong homogeneous magnetic field. For the sake of simplicity we
work in the extreme relativistic limit, where we neglect the masses
of electrons as well as muons. In the presence of magnetic field
the form of spinor and hence the form of electron propagator is 
changed
but the form of photon propagator remains the same.  So first we are
going to revisit the Dirac equation in the strong magnetic field to 
obtain
the form of positive energy and negative energy spinors and their 
completeness relations.

\subsection{Notations}
\begin{figure}[h]
	\begin{center}
	\includegraphics[height=12cm,width=8cm,angle=-90]{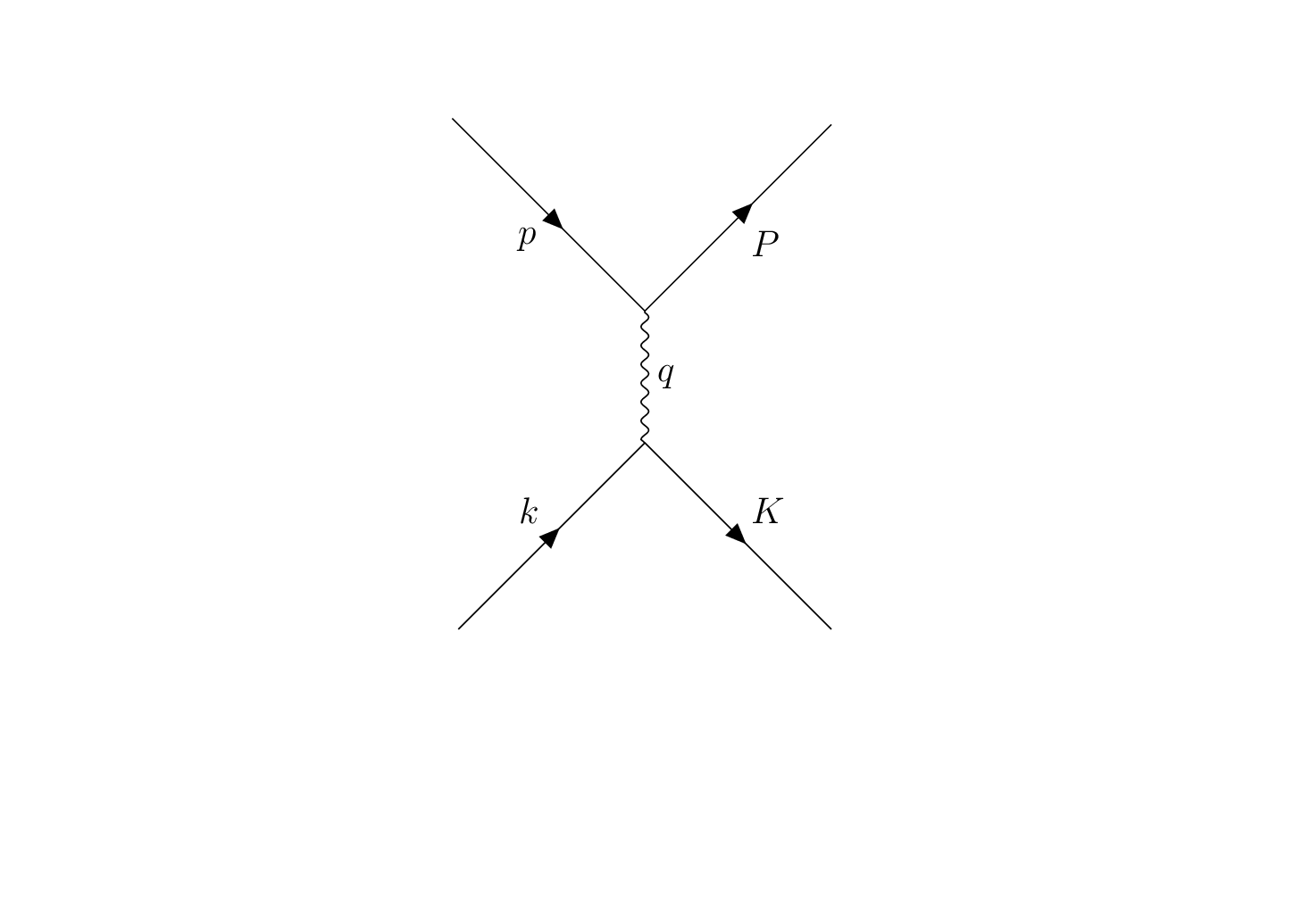}
	\vspace{-2cm}
	\caption{}\label{notation}
	\end{center}
\end{figure}
The dynamics of an electron in a magnetic field is factorized into 
transverse and longitudinal plane with respect to the direction of 
magnetic 
field, as a result its momentum ($\mathbf{p}$) is separated into 
perpendicular and longitudinal components with respect to the direction 
of magnetic field
(say, $z$-direction). Hence the dispersion relation is modified 
quantum mechanically into
\begin{eqnarray}
E_n(p_z)=\sqrt{p_z^2+m_f^2+2n\left|eB\right|}\quad,\label{landau_quantization}
\end{eqnarray}
In a strong magnetic field ($|eB|>>m^2$), electrons 
prefer to lie in the lowest Landau level, hence the electron 
momentum becomes
purely longitudinal~\cite{Gusynin:1998nh}, i.e $\vec{p}_\perp \approx 0$.
The aforesaid observation motivates
to construct the following kinematic variables, {\em viz.}
momentum, Mandelstam variables etc. which will be advantageous
to express matrix element squared, crosssection etc. in a strong magnetic 
field. Thus using the following convention of the metric tensor
\begin{eqnarray}
g^{\mu\nu}=(1,-1,-1,-1), g^{\mu_{\perp}\nu_{\perp}}=(0,-1,-1,0)~ {\rm and} ~ 
g^{\mu_{\parallel}\nu_{\parallel}}=(1,0,0,-1)~,
\end{eqnarray}
we will first denote the four-momentum for a generic Feynman diagram in 
Figure \ref{notation},
\begin{eqnarray}
p_\perp^\mu&=&(0,p^1,p^2,0)=(0,p_x,p_y,0),\\
p_\parallel^\mu&=&(p^0,0,0,p^3)=(E_0,0,0,p_z), \\
\widetilde{p_\parallel}^\mu&=&(\widetilde{p}^0,0,0,\widetilde{p}^3)
=(p^3,0,0,p^0),
\end{eqnarray}
therefore the usual Mandelstam variables take the following form: 
\begin{eqnarray}
s&=&(p_\parallel+k_\parallel)^2=(P_\parallel+K_\parallel)^2,\label{mandelstam-s}\\
t&=&(p_\parallel-P_\parallel)^2=(K_\parallel-k_\parallel)^2,\label{mandelstam-t}\\
u&=&(p_\parallel-K_\parallel)^2=(P_\parallel-k_\parallel)^2.\label{mandelstam-u}
\end{eqnarray}
 We define some new variables 
 $s_p,s_k,s_P,s_K,t_p,t_P,t_K,t_k,u_p,u_k,u_P,u_K$, dubbed as the magnetic 
 Mandelstam variables, which are defined as
 \begin{eqnarray}
 s_p&=&(\widetilde p_\parallel+k_\parallel)^2,~s_k=(\widetilde k_\parallel+p_\parallel)^2,\label{magnetic-mandelstam-s}\\
 s_P&=&(\widetilde P_\parallel+K_\parallel)^2,~s_K=(\widetilde K_\parallel+P_\parallel)^2,\label{magnetic-mandelstam-ss}\\
 t_p&=&(\widetilde p_\parallel-P_\parallel)^2,~t_P=(\widetilde P_\parallel-p_\parallel)^2,\label{magnetic-mandelstam-t}\\
 t_K&=&(\widetilde K_\parallel-k_\parallel)^2,~t_k=(\widetilde k_\parallel-K_\parallel)^2,\label{magnetic-mandelstam-tt}\\
 u_p&=&(\widetilde p_\parallel-K_\parallel)^2,~u_K=(\widetilde K_\parallel-p_\parallel)^2,\label{magnetic-mandelstam-u}\\
 u_P&=&(\widetilde P_\parallel-k_\parallel)^2,~u_k=(\widetilde k_\parallel-P_\parallel)^2.\label{magnetic-mandelstam-uu}
 \end{eqnarray}
Although the form of the magnetic Mandelstam variables (or the 
half-tilde
Mandelstam variables, where one momentum is tilde) are same as the 
Mandelstam
variables but they are completely different from them. In the 
extreme 
relativistic limit, they satisfy the following relations among 
themselves
\begin{eqnarray}
	s_p=-s_k,&&s_P=-s_K,\\
	t_p=-t_P,&&t_K=-t_k,\\
	u_p=-u_K,&&u_P=-u_k.
\end{eqnarray}
Furthermore we use other notations for the full-tilde Mandelstam 
variables,
which are defined as 
\begin{eqnarray}
	\widetilde s&=&(\widetilde p_\parallel+\widetilde k_\parallel)^2=(\widetilde P_\parallel+\widetilde K_\parallel)^2,\\
	\widetilde t&=&(\widetilde p_\parallel-\widetilde P_\parallel)^2=(\widetilde K_\parallel-\widetilde k_\parallel)^2,\\
	\widetilde u&=&(\widetilde p_\parallel-\widetilde K_\parallel)^2=(\widetilde P_\parallel-\widetilde k_\parallel)^2.
\end{eqnarray}
We can directly relate the full-tilde Mandelstam variables to the 
Mandelstam variables as
\begin{eqnarray}
	s=-\widetilde s,~t=-\widetilde t,~u=-\widetilde u~.
\end{eqnarray}

\subsection{Dirac Spinors in a strong magnetic field}
The methods of Ritus eigenfunction~\cite{ritus} along with 
the Schwinger Proper-time formalism~\cite{schwinger} are commonly 
used to 
solved the Dirac equation of charged fermions in the presence of a 
constant magnetic field. There are different ways which have been 
adopted in the literature~\cite{p_meszaros,herold_ruder_wunner,
sokolov_ternov} to obtain the spinor in a magnetic field, however,
we have mainly adopted to solve the 
Dirac equation in a constant external field 
from Ref.~\cite{Bhattacharya:2007vz}. 

For the sake of simplicity, we assume a static and 
homogeneous magnetic field, which is along the $z$-direction,
$\vec{B} =B \hat{z}$.  Such a magnetic field can be
obtained from a vector potential $A^\mu = (0,0,Bx,0)$.
The choice of vector potential is not unique as the same magnetic
field can also be obtained from a symmetric potential given by
$A^\mu = (0,\frac{-By}{2},\frac{Bx}{2},0)$. Thus the positive energy 
Dirac spinors with the gauge $A^\mu=(0,-By,0,0)$ are given by
the shifted coordinate, $\xi$ (=$\sqrt{eB}\left(y-\frac{p_x}{eB}\right)$)
~\cite{Furry,Bhattacharya:2007vz}
 	\begin{eqnarray}
 	U_+ (y,n,\vec p_{\omit{y}})=N\left( \begin{array}{c} I_{n-1}(\xi) \\[2ex] 0 \\[2ex] 
 	{\strut\textstyle p_z \over \strut\textstyle E_n+m} I_{n-1}(\xi) \\[2ex]
 	-\, {\strut\textstyle \sqrt{2neB} \over \strut\textstyle 
 		E_n+m} I_n (\xi) 
 	\end{array} \right);\hspace{2mm}
 	U_- (y,n,\vec p_{\omit{y}})=N\left( \begin{array}{c} 0 \\[2ex] I_n (\xi) \\[2ex]
 	-\, {\strut\textstyle \sqrt{2neB} \over \strut\textstyle E_n+m}
 	I_{n-1}(\xi) \\[2ex] 
 	-\,{\strut\textstyle p_z \over \strut\textstyle E_n+m} I_n(\xi) 
 	\end{array} \right)
 	\end{eqnarray}
Similarly the negative energy Dirac spinors with $\widetilde \xi$ 
(=$\sqrt{eB}\left(y+\frac{p_x}{eB}\right)$) are given 
by~\cite{Furry,Bhattacharya:2007vz}
 	\begin{eqnarray}
 	V_- (y,n,\vec p_{\omit{y}}) = N\left( \begin{array}{c} 
 	{\strut\textstyle p_z \over \strut\textstyle E_n+m}
 	I_{n-1}(\widetilde\xi) \\[2ex] 
 	{\strut\textstyle \sqrt{2neB} \over \strut\textstyle E_n+m} 
 	I_n (\widetilde\xi)  \\[2ex] 
 	I_{n-1}(\widetilde\xi) \\[2ex] 0
 	\end{array} \right);\hspace{2mm}
 	V_+ (y,n,\vec p_{\omit{y}}) =N \left( \begin{array}{c} 
 	{\strut\textstyle \sqrt{2neB} \over \strut\textstyle E_n+m}
 	I_{n-1}(\widetilde\xi) \\[2ex] 
 	-\,{\strut\textstyle p_z \over \strut\textstyle E_n+m}
 	I_n(\widetilde\xi)  \\[2ex] 
 	0 \\[2ex] I_n (\widetilde\xi)
 	\end{array} \right) \,
 	\end{eqnarray}
where the normalization constant ($N$) is $N=\sqrt{E_n+m}$ and the 
symbol,
$p_{\omit y}$ denotes the absence of the $y$-component of momentum 
in the
spinors. The energy eigenvalues are given by the above Landau 
quantization
\eqref{landau_quantization},
where $n$ denotes the Landau levels and the energy eigenfunctions, 
$I_n (\xi)$ 
are expressed in terms of Hermite polynomials, $H_n (\xi)$
 	\begin{eqnarray}
 	I_n(\xi)&=&\frac{\sqrt{eB}}{n!2^n\sqrt{\pi}}e^{\frac{-\xi^2}{2}}H_n(\xi),
 	\end{eqnarray}
with the properties: $I_{-1}(\xi)=0$ and $I_0^2(\xi)=1$.
As mentioned earlier, in a strong magnetic field, only the lowest
Landau level (n=0) is populated.
 	
We can now calculate the spin sums for the particles ($P_U$) and 
anti-particles ($P_V$) in the presence of external magnetic field  
as~\cite{Furry,Bhattacharya:2007vz}
 	    \begin{eqnarray}
 	P_U (y,y' ,n,\vec p\omit y) &=&\sum_{\rm s}U_s (y,n,\vec p_{\omit{y}})
\overline{U}_s (y',n,\vec p_{\omit{y}})\nonumber\\
&=& {1\over 2} \bigg[\left\{ m(1+\Sigma_z) +
 	\rlap/p_\parallel - 
 	\widetilde{\rlap/p}_\parallel \gamma_5 \right\} I_{n-1}(\xi)
 	I_{n-1}(\xi')\nonumber\\
 	&+& \left\{ m(1-\Sigma_z) + \rlap/p_\parallel -
 	\gamma_5 \widetilde{\rlap/p}_\parallel  \right\} I_n(\xi)
 	I_n (\xi')\nonumber\\ 
 	&-& \sqrt{2neB} (\gamma_1 - i\gamma_2) I_n(\xi) I_{n-1}(\xi') \nonumber\\
 	&-& \sqrt{2neB} (\gamma_1 + i\gamma_2) I_{n-1}(\xi) I_n(\xi') 
 	\bigg],\label{PUs} 
 	\end{eqnarray}
 	\begin{eqnarray}
 	P_V(\widetilde y,\widetilde{y'} ,n,\vec p\omit y) &=&\sum_{\rm s}V_s (\widetilde{y},n,
\vec p_{\omit{y}})\overline{V}_s (\widetilde{y'},n,\vec p_{\omit{y}})
\nonumber\\
&=& {1\over 2}  \Bigg[ \left\{ -m(1+\Sigma_z) +
 	\rlap/p_\parallel - 
 	\widetilde{\rlap/p}_\parallel \gamma_5 \right\} I_{n-1}(\widetilde\xi)
 	I_{n-1} (\widetilde\xi')\nonumber\\ 
 	&+& \left\{ -m(1-\Sigma_z) + \rlap/p_\parallel -
 	\gamma_5 \widetilde{\rlap/p}_\parallel  \right\} I_n(\widetilde\xi)
 	I_n(\widetilde\xi')\nonumber\\ 
 	&+&\sqrt{2neB} (\gamma_1 - i\gamma_2) I_n(\widetilde\xi)
 	I_{n-1}(\widetilde\xi')\nonumber\\
 	&+& \sqrt{2neB} (\gamma_1 + i\gamma_2) I_{n-1}(\widetilde\xi)
 	I_n(\widetilde\xi') \Bigg], \,
 	\label{PVs}
 	\end{eqnarray}
where $\Sigma_z=i\gamma^1\gamma^2$, 
$\slashed{p}_\parallel=p^0\gamma^0-p^3\gamma^3$, 
$\widetilde{\slashed{p}}_\parallel=p^3\gamma^0-p^0\gamma^3$ and $m$ 
is the mass of electron. 

\subsection{Matrix element for electron-muon scattering at the lowest-order}
There are two different processes for electron-muon 
scattering, {\em viz.} $\ems$ and $\emt$ represented by $s$ and $t$-channel
diagrams, respectively.
We will first evaluate the matrix element ($\mathfrak {M}$) for the 
$s$-channel diagram and calculate the $\overline{\left|\mathfrak{M}\right|^2}$
by summing over the spin states. Finally we calculate the 
same for the $t$-channel process.\\

\begin{figure}[h]
	\begin{center}
	\includegraphics[height=12cm,width=8cm,angle=-90]{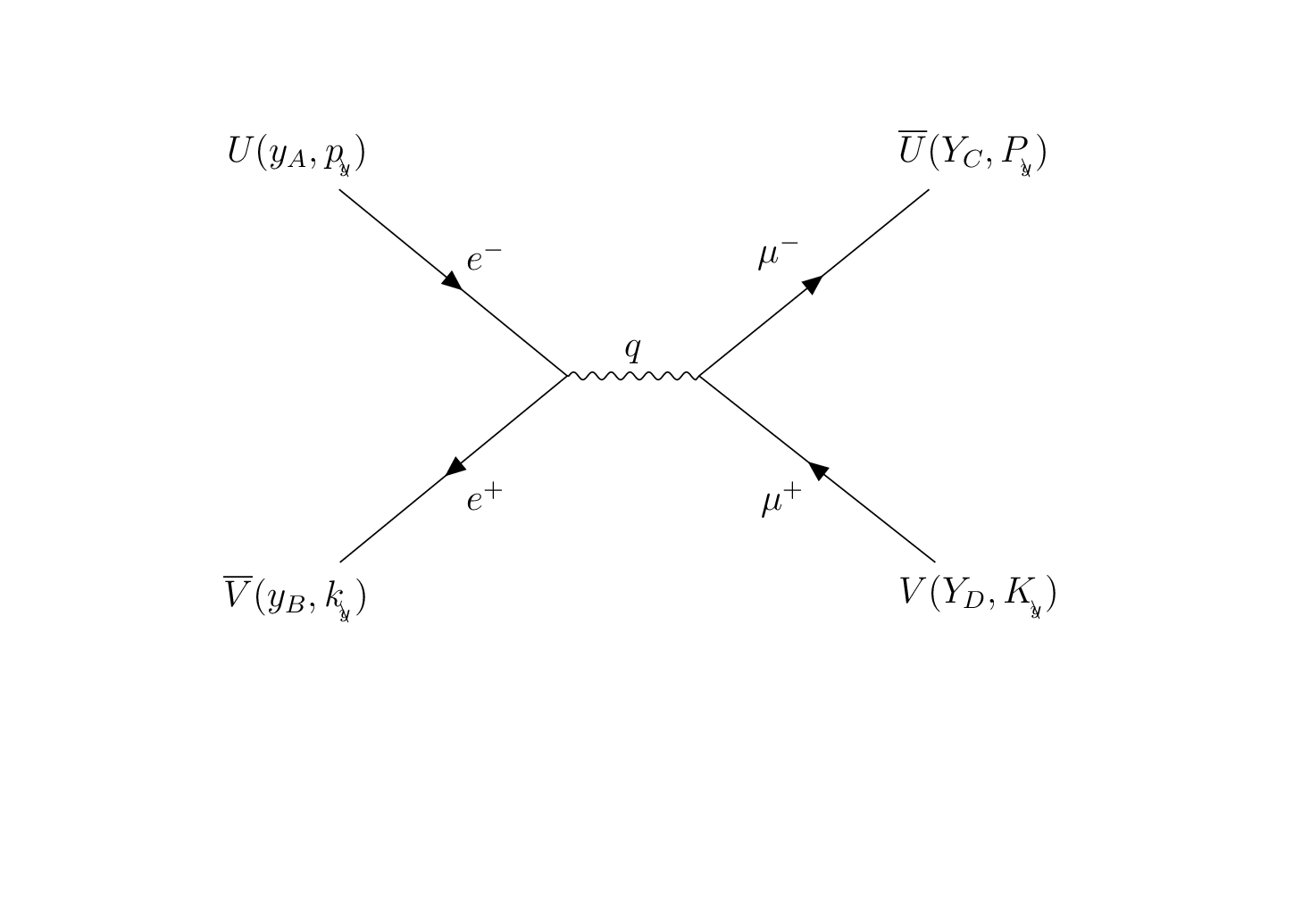}
	\vspace{-2cm}
	\caption{$s$-channel}\label{s_e_mu}
	\end{center}
\end{figure}

\noindent {\large \bf $s$-Channel Process:}\\
The electron-muon scattering in $s$-channel represents the $e^-  e^+ 
\rightarrow \mu^-  \mu^+$ process by the following Feynman diagram
at the lowest-order in Figure \ref{s_e_mu}.
where $U$ and $\overline U$ denote the Dirac spinors for the 
incoming and the outgoing fermions, 
respectively whereas $V$ and $\overline V$ in Figure \ref{s_e_mu} 
represent the spinors for the outgoing and the incoming 
anti-fermions,
respectively. Therefore the invariant amplitude for the $e^-  e^+ 
\rightarrow \mu^-  \mu^+$ process at the lowest-order
is given by the matrix element,
\begin{eqnarray}
	-i\mathfrak{M}_{\rm s} \ems &=&\big[\overline{V}(y_B,k_{\omit{y}})ie\gamma^\mu U(y_A,p_{\omit{y}})\big]
\left(\frac{-ig^{\mu\nu}}{q^2}\right)\big[\overline{U}(Y_C,P_{\omit{y}})ie\gamma^\nu 
V(Y_D,K_{\omit{y}})\big].\label{s-matrix}
\end{eqnarray}

We will now calculate the square of the matrix element, 
${\left|\mathfrak{M}_s\right|}^2$ 
and then sum over the spin-states. It is convenient and easy to 
solve if we 
separately write the spin sums for electron and muon
\begin{equation}
	\overline{\left|\mathfrak{M}_{\rm s}\right|^2}{\ems}=\frac{e^4}{q^4}L_{e}^{\mu\nu}
L^{muon}_{\mu\nu},\label{matrix-element-square}
\end{equation}
where the $L_e^{\mu \nu}$ and $L^{\rm{muon}}_{\mu \nu}$ are given by
\begin{eqnarray}
	L_{e}^{\mu\nu}&=&\sum_{e~spins}\big[\overline{V}(y_B,k_{\omit{y}})
\gamma^\mu U(y_A,p_{\omit{y}})\big]\big[\overline{V}(y_B,k_{\omit{y}})\gamma^\nu U(y_A,p_{\omit{y}})
\big]^*~,\\
L^{muon}_{\mu\nu}&=&\sum_{\mu~ spins}\big[\overline{U}(Y_C,P_{\omit{y}})\gamma_\mu 
V(Y_D,K_{\omit{y}})\big]\big[\overline{U}(Y_C,P_{\omit{y}})\gamma_\nu V(Y_D,K_{\omit{y}})\big]^*~,
\end{eqnarray}
respectively. 

Using the properties of gamma matrices, {\em like}
\begin{equation*}
	{\gamma^\mu}^\dag=\gamma^0\gamma^\mu\gamma^0 \hspace{3mm}\mbox{and}\hspace{3mm}\gamma^0\gamma^0=I,
\end{equation*}
and the cyclic property of trace, we further simplify $L_{e}^{\mu\nu}$ as
\begin{eqnarray}
L_{e}^{\mu\nu}=Tr\big[P_V(y_B,k_{\omit{y}})\gamma^\mu P_U(y_A,p_{\omit{y}})\gamma^\nu\big]
~.
\end{eqnarray}
The above spin-sums, $P_V$ and $P_U$ for the positron and the electron 
~\cite{Furry,Bhattacharya:2007vz} in strong 
magnetic field ({\em i.e.} for the lowest Landau levels, $n=0$) can 
be calculated as 
\begin{eqnarray}
	P_V(y_B,n=0,k_\parallel)&=&\frac{1}{2}I_0^2(\xi_B)\big[-m(1-\Sigma_z)+
\slashed{k}_\parallel-\gamma_5\widetilde{\slashed{k}}_\parallel\big],\\
	P_U(y_A,n=0,p_\parallel)&=&\frac{1}{2}I_0^2(\xi_A)\big[m(1-\Sigma_z)+
\slashed{p}_\parallel-\gamma_5\widetilde{\slashed{p}}_\parallel\big],\\
\end{eqnarray}
respectively.

Thus, after calculating the traces \footnote{which 
is calculated in Appendix A}, the tensor at the electron vertex 
is simplified into
\begin{equation}\label{L-electron}
L_{e}^{\mu\nu}=\Big[p^\mu_\parallel k^\nu_\parallel+p^\nu_\parallel k^\mu_\parallel-(p_\parallel\cdot k_\parallel) g^{\mu\nu}
+\widetilde p^\mu_\parallel \widetilde k^\nu_\parallel+\widetilde p^\nu_\parallel \widetilde k^\mu_\parallel-
(\widetilde p_\parallel\cdot \widetilde k_\parallel) g^{\mu\nu}\Big]-2m^2(g^{\mu\nu}-g^{\mu_{\perp}\nu_{\perp}}).
\end{equation}
In a similar way the tensor at the muon vertex is also calculated as
\begin{equation}\label{L-muon}
L^{muon}_{\mu\nu}=\Big[K_{\parallel \mu} P_{\parallel \nu}+K_{\parallel \nu} P_{\parallel \mu}-(K_\parallel\cdot P_\parallel) g_{\mu\nu}+
\widetilde K_{\parallel \mu} \widetilde P_{\parallel \nu}+\widetilde K_{\parallel \nu} \widetilde P_{\parallel \mu}-
(\widetilde K_\parallel\cdot \widetilde P_\parallel) g_{\mu\nu}\Big]-
2M^2(g_{\mu\nu}-g_{\mu_{\perp}\nu_{\perp}}).
\end{equation}
However in the extreme relativistic limit, where the mass terms 
could be
neglected, the above squared matrix element 
(\ref{matrix-element-square})
becomes simplified and is given by the short-hand notation
\begin{equation}\label{matrix-element-squre-2}
\overline{\left| \mathfrak{M}_{\rm s} \right|^2}{\ems}=\frac{e^4}{q^4}(T^{\mu\nu}R_{\mu\nu}+T^{\mu\nu}\widetilde{R}_{\mu\nu}+\widetilde{T}^{\mu\nu}R_{\mu\nu}
+\widetilde{T}^{\mu\nu}\widetilde{R}_{\mu\nu})~,
\end{equation}
where $T^{\mu\nu},R_{\mu\nu},\widetilde{T}^{\mu\nu},\widetilde{R}_{\mu\nu}$ are defined by
\begin{eqnarray}
T^{\mu\nu}&=&p^\mu_\parallel k^\nu_\parallel+p^\nu_\parallel k^\mu_\parallel-(p_\parallel\cdot k_\parallel) g^{\mu\nu},\\
\widetilde{T}^{\mu\nu}&=&\widetilde p^\mu_\parallel \widetilde k^\nu_\parallel+\widetilde p^\nu_\parallel 
\widetilde k^\mu_\parallel-(\widetilde p_\parallel\cdot \widetilde k_\parallel) g^{\mu\nu},\\
R_{\mu\nu}&=&K_{\parallel \mu} P_{\parallel \nu}+K_{\parallel \nu} P_{\parallel \mu}-(K_\parallel\cdot P_\parallel) g_{\mu\nu},\\
\widetilde{R}_{\mu\nu}&=&\widetilde K_{\parallel \mu} \widetilde P_{\parallel \nu}+\widetilde K_{\parallel \nu} \widetilde P_{\parallel \mu}-
(\widetilde K_\parallel\cdot \widetilde P_\parallel) g_{\mu\nu}.
\end{eqnarray}
Thus the products are calculated as
\begin{eqnarray}
	T^{\mu\nu}R_{\mu\nu}&=&2\Big[(p_\parallel\cdot K_\parallel)(k_\parallel\cdot P_\parallel)+(p_\parallel\cdot P_\parallel)(k_\parallel\cdot K_\parallel)\Big],\\
	\widetilde{T}^{\mu\nu}R_{\mu\nu}&=&2\Big[(\widetilde p_\parallel\cdot K_\parallel)(\widetilde k_\parallel\cdot P_\parallel)+(\widetilde 
p_\parallel\cdot P_\parallel)(\widetilde k_\parallel\cdot K_\parallel)\Big],\\
	T^{\mu\nu}\widetilde{R}_{\mu\nu}&=&2\Big[(p_\parallel\cdot \widetilde K_\parallel)(k_\parallel\cdot \widetilde P_\parallel)+(p_\parallel\cdot 
\widetilde P_\parallel)(k_\parallel\cdot \widetilde K_\parallel)\Big],\\
	\widetilde{T}^{\mu\nu}\widetilde{R}_{\mu\nu}&=&2\Big[(\widetilde p_\parallel\cdot \widetilde K_\parallel)(\widetilde k_\parallel\cdot 
\widetilde P_\parallel)+(\widetilde p_\parallel\cdot \widetilde P_\parallel)(\widetilde k_\parallel\cdot 
\widetilde K_\parallel)\Big].
\end{eqnarray}

Thus after substituting the products, the matrix element squared 
\eqref{matrix-element-squre-2} 
for the electron-muon scattering for $s$-channel becomes
\begin{eqnarray}
	\overline{\left|\mathfrak{M}_{\rm s}\right|^2}{\ems}&=&\frac{2e^4}{q^4}\Big[(p_\parallel\cdot K_\parallel)(k_\parallel\cdot P_\parallel)+(p_\parallel\cdot P_\parallel)(k_\parallel\cdot K_\parallel)+(p_\parallel\cdot \widetilde K_\parallel)(k_\parallel\cdot \widetilde P_\parallel)\nonumber\\
	&&+(p_\parallel\cdot \widetilde P_\parallel)(k_\parallel\cdot \widetilde K_\parallel)+(\widetilde p_\parallel\cdot K_\parallel)(\widetilde k_\parallel\cdot P_\parallel)+(\widetilde p_\parallel\cdot P_\parallel)(\widetilde k_\parallel\cdot K_\parallel)\nonumber\\
	&&+(\widetilde p_\parallel\cdot \widetilde K_\parallel)(\widetilde k_\parallel\cdot \widetilde P_\parallel)+(\widetilde p_\parallel\cdot \widetilde P_\parallel)(\widetilde k_\parallel\cdot \widetilde K_\parallel)\Big].  \label{matrix-element-square-3}
\end{eqnarray}

Using the notations for the Mandelstam and magnetic Mandelstam 
variables
mentioned in equations \eqref{mandelstam-s}-\eqref{mandelstam-u} 
and 
\eqref{magnetic-mandelstam-s}-\eqref{magnetic-mandelstam-uu}, 
respectively, 
the above matrix element squared becomes 
\begin{eqnarray}
{\overline{\left|\mathfrak{M}_{\rm s}\right|^2}}_{\rm B \neq 0} (e^- e^+ \rightarrow \mu^- 
\mu^+)&=&
\frac{e^4}{2q^4}\Big[u^2+t^2+u_Ku_P+t_Pt_K+u_pu_k+t_pt_k+\widetilde{u}^2+\widetilde{t}^2\Big]\nonumber\\
&=&\frac{e^4}{s^2}\Big[u^2+t^2+u_pu_k+t_pt_k\Big].\label{matrix-element-square-4}
\end{eqnarray}

For the sake of completeness, the $e$-$\mu$ scattering in 
$s$-channel, 
{\em i.e.} $\ems$ process in vacuum can be calculated 
as~\cite{halzen_martin}
\begin{eqnarray}
{\overline{\left|\mathfrak{M}_{\rm s}\right|^2}}_{\rm B =0} \ems=
\frac{2e^4}{s^2}\Big[u^2+t^2\Big].\label{s-vacuum}
\end{eqnarray}

\begin{figure}[h]
	\begin{center}
	\includegraphics[height=12cm,width=8cm,angle=-90]{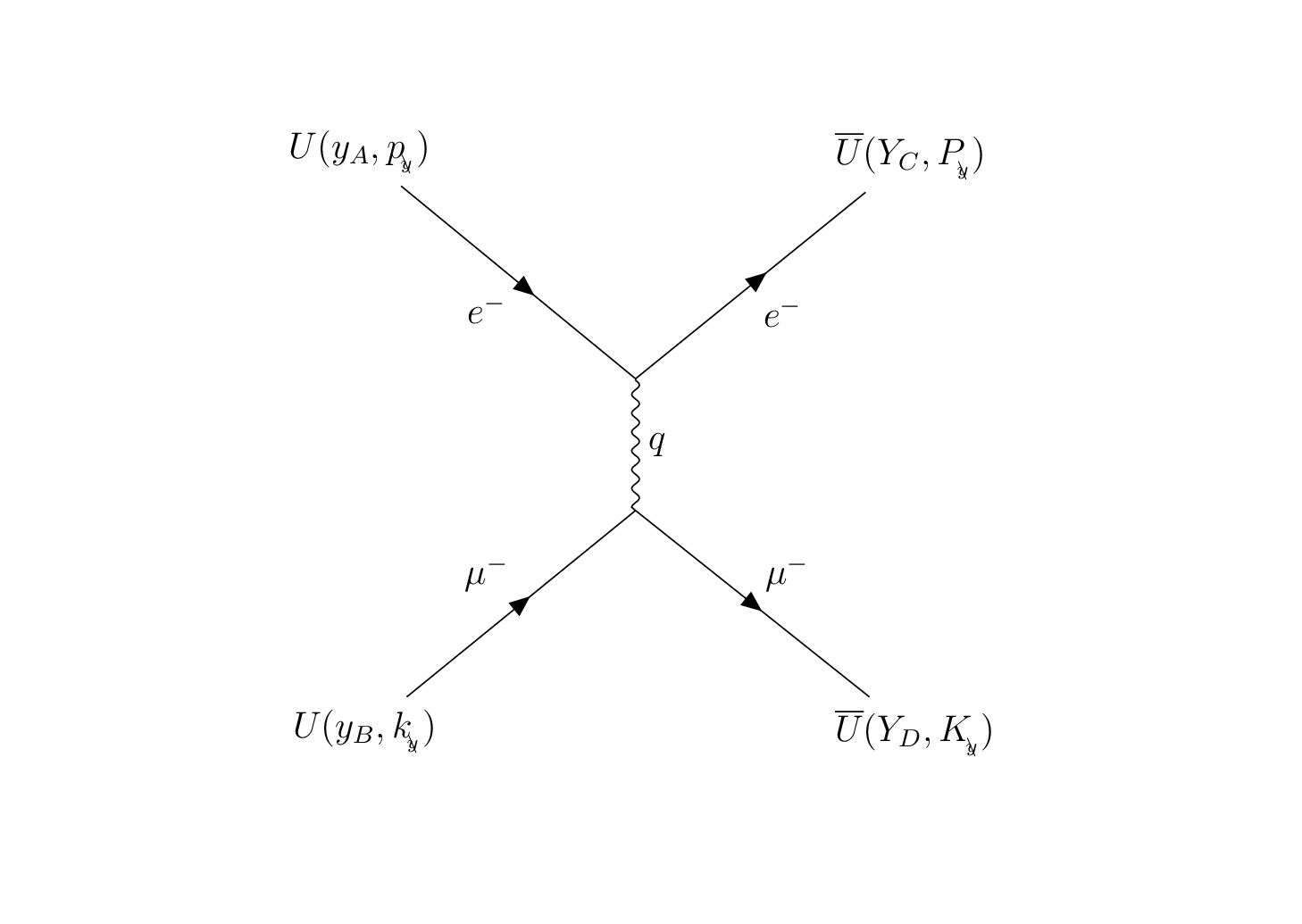}
	\vspace{-1cm}
	\caption{$t$-channel}\label{t_e_mu}
	\end{center}
\end{figure}

\noindent {\large \bf $t$-Channel Process:}\\
The electron-muon scattering in the $t$-channel, {\em i.e.} $e^-  \mu^- 
\rightarrow e^- \mu^-$ process in the lowest-order is represented 
by the following Feynman diagram in Figure \ref{t_e_mu}.
We will now evaluate the matrix element for it as
\begin{eqnarray}
\mathfrak{M}_{\rm t} \emt &=& \frac{-e^2}{q^2}\big[\overline{U}(Y_C,P_{\omit{y}})
\gamma^\mu U(y_A,p_{\omit{y}})\big]\big[\overline{V}(y_B,k_{\omit{y}})
\gamma_\mu V(Y_D,K_{\omit{y}})\big].
\end{eqnarray}

As we know that the $e$-$\mu$  scattering in $s$-channel, {\em i.e.} 
$e^-  e^+ \rightarrow \mu^-  \mu^+$ process is related to
the $e$-$\mu$  scattering in $t$-channel, {\em i.e.} 
$e^-  \mu^- \rightarrow e^-\mu^- $ process by the
crossing symmetry. This facilitates us to obtain the spin-summed 
squared matrix element for the $t$-channel directly from the 
$e$-$\mu$ scattering in $s$-channel in following way.

As mentioned earlier, in the strong magnetic field (along the
$z$-direction), the spinors do not have the spatial $y$ dependence
and the perpendicular component of momentum ($p_\perp$) becomes
vanishingly small. Therefore the matrix element in $s$-channel \eqref{s-matrix}
can be rewritten as
	\begin{eqnarray}
	\mathfrak{M}_{\rm s} \ems &=&\frac{-e^2}{(p_\parallel+k_\parallel)^2}\big[\overline{V}(k_\parallel)\gamma^\mu U(p_\parallel)\big]
	\big[\overline{U}(P_\parallel)\gamma_\mu 
	V(K_\parallel)\big].
	\end{eqnarray}
Now if we interchange the momentum $k_\parallel$ with $-P_\parallel$ due to
the crossing symmetry, the above matrix element becomes
	\begin{eqnarray}
	\mathfrak{M}_{\rm s} \ems &=&\frac{-e^2}{(p_\parallel-P_\parallel)^2}\big[\overline{V}(-P_\parallel)\gamma^\mu U(p_\parallel)\big]
	\big[\overline{U}(-k_\parallel)\gamma_\mu 
	V(K_\parallel)\big].
	\end{eqnarray}
Using the Feynman-St\"uckelberg interpretation, where by reversing the 
direction of  momentum (say, p) in the spinors, $\overline{V}(-p)$ becomes 
$\overline{U}(p)$ and $\overline{U}(-p)$ becomes $\overline{V}(p)$, the above $s$-channel matrix element gives the desired 
$t$-channel matrix element 
	\begin{eqnarray}
	\mathfrak{M}_{\rm t} \emt &=& \frac{-e^2}{(p_\parallel-P_\parallel)^2}\big[\overline{U}(P_\parallel)
	\gamma^\mu U(p_\parallel)\big]\big[\overline{V}(k_\parallel)
	\gamma_\mu V(K_\parallel)\big].
	\end{eqnarray}
Therefore the spin-summed matrix element squared for the $t$-channel diagram
(Figure 3) is easily derived as
\begin{eqnarray}
\overline{\left|\mathfrak{M}_{\rm t}\right|^2}_{\rm B \neq 0} {\emt}&=&\frac{2e^4}{q^4}\Big[(p_\parallel\cdot k_\parallel)(P_\parallel\cdot K_\parallel)+(p_\parallel\cdot K_\parallel)(P_\parallel\cdot k_\parallel)+(p_\parallel\cdot \widetilde k_\parallel)(P_\parallel\cdot \widetilde K_\parallel)\nonumber\\
&&+(p_\parallel\cdot \widetilde K_\parallel)(P_\parallel\cdot \widetilde k_\parallel)+(\widetilde p_\parallel\cdot k_\parallel)(\widetilde P_\parallel\cdot K_\parallel)+(\widetilde p_\parallel\cdot K_\parallel)(\widetilde P_\parallel\cdot k_\parallel)\nonumber\\
&&+(\widetilde p_\parallel\cdot \widetilde k_\parallel)(\widetilde P_\parallel\cdot \widetilde K_\parallel)
+(\widetilde p_\parallel\cdot \widetilde K_\parallel)(\widetilde P_\parallel
\cdot \widetilde k_\parallel)\Big],\label{e-u-t-ch}
\end{eqnarray}
which, in turn, will be expressed in terms of the Mandelstam 
variables from the 
$s$-channel expression \eqref{matrix-element-square-4} by 
interchanging 
$s$ with $t$, $u_k$ with $u_P$, $s_k$ with $t_P$, and $s_P$ with $t_k$ and is 
given by
\begin{eqnarray}
{\overline{\left|\mathfrak{M}_{\rm t}\right|^2}}_{\rm B \neq 0} {(e^- \mu^- 
\rightarrow e^- \mu^-)}&=&
\frac{e^4}{4q^4}\Big[s^2+u^2+s_ks_K+u_Ku_k+s_ps_P+u_pu_P+\widetilde{s}^2+\widetilde{u}^2\Big]\nonumber\\
&=&\frac{e^4}{t^2}\Big[s^2+u^2+s_ps_P+u_pu_P\Big].\label{matrix-element-squared-t}
\end{eqnarray}
However, the above squared-matrix element in vacuum 
is~\cite{halzen_martin}
\begin{eqnarray}
{\overline{\left|\mathfrak{M}_{\rm t}\right|^2}}_{\rm B=0} \emt=
\frac{2e^4}{t^2}\Big[s^2+u^2\Big].\label{t-vacuum}
\end{eqnarray}
\subsection{Matrix Element for Bhabha Scattering: $e^- e^+ \rightarrow 
e^- e^+$}
\begin{figure}[h]
\begin{minipage}{0.5\textwidth}
		\includegraphics[height=10cm,width=8cm,angle=-90]{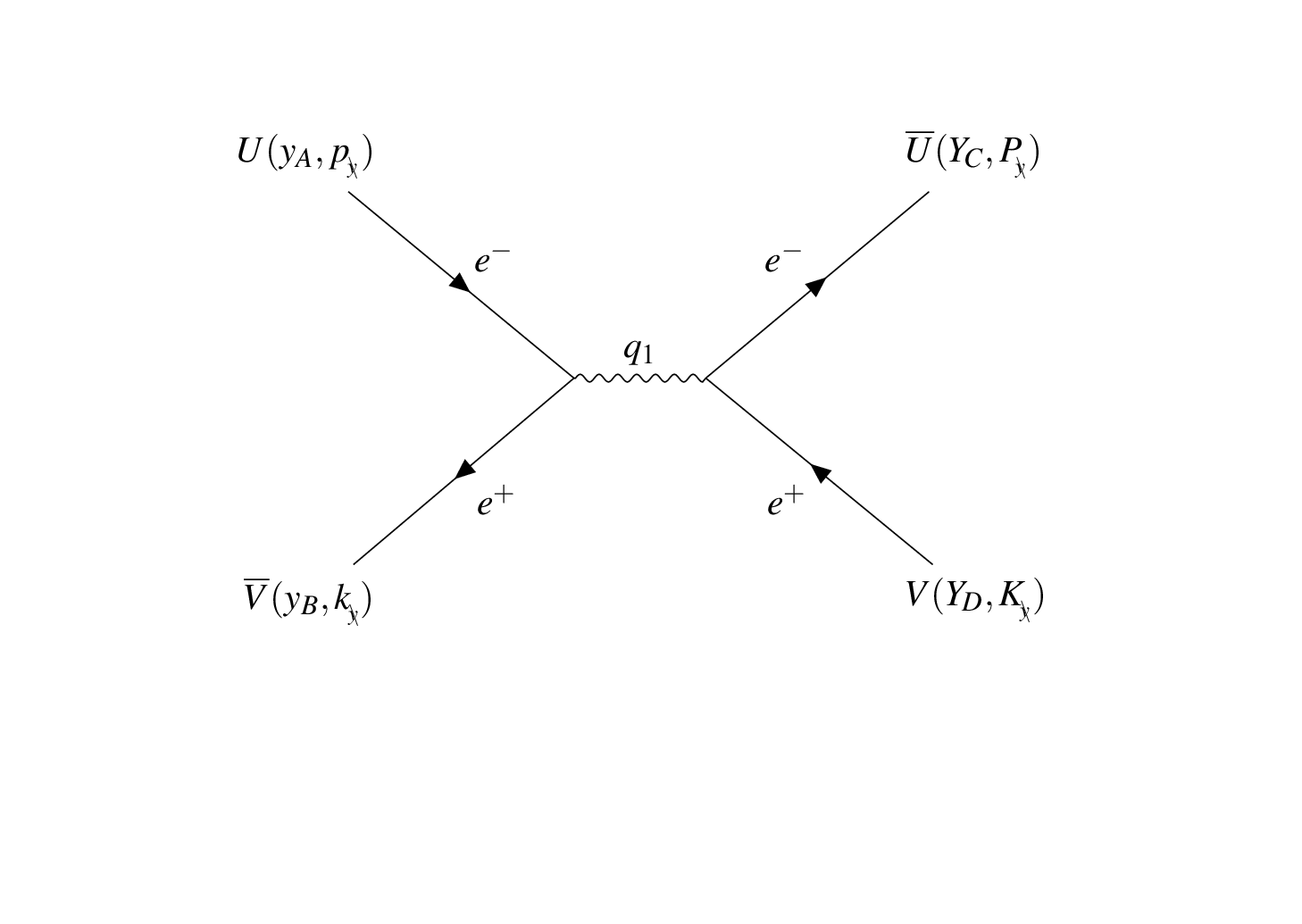}
		\vspace{-2cm}
	\caption{$s$-channel}\label{s-ch}
\end{minipage}
\begin{minipage}{0.5\textwidth}
		\includegraphics[height=9cm,width=7cm,angle=-90]{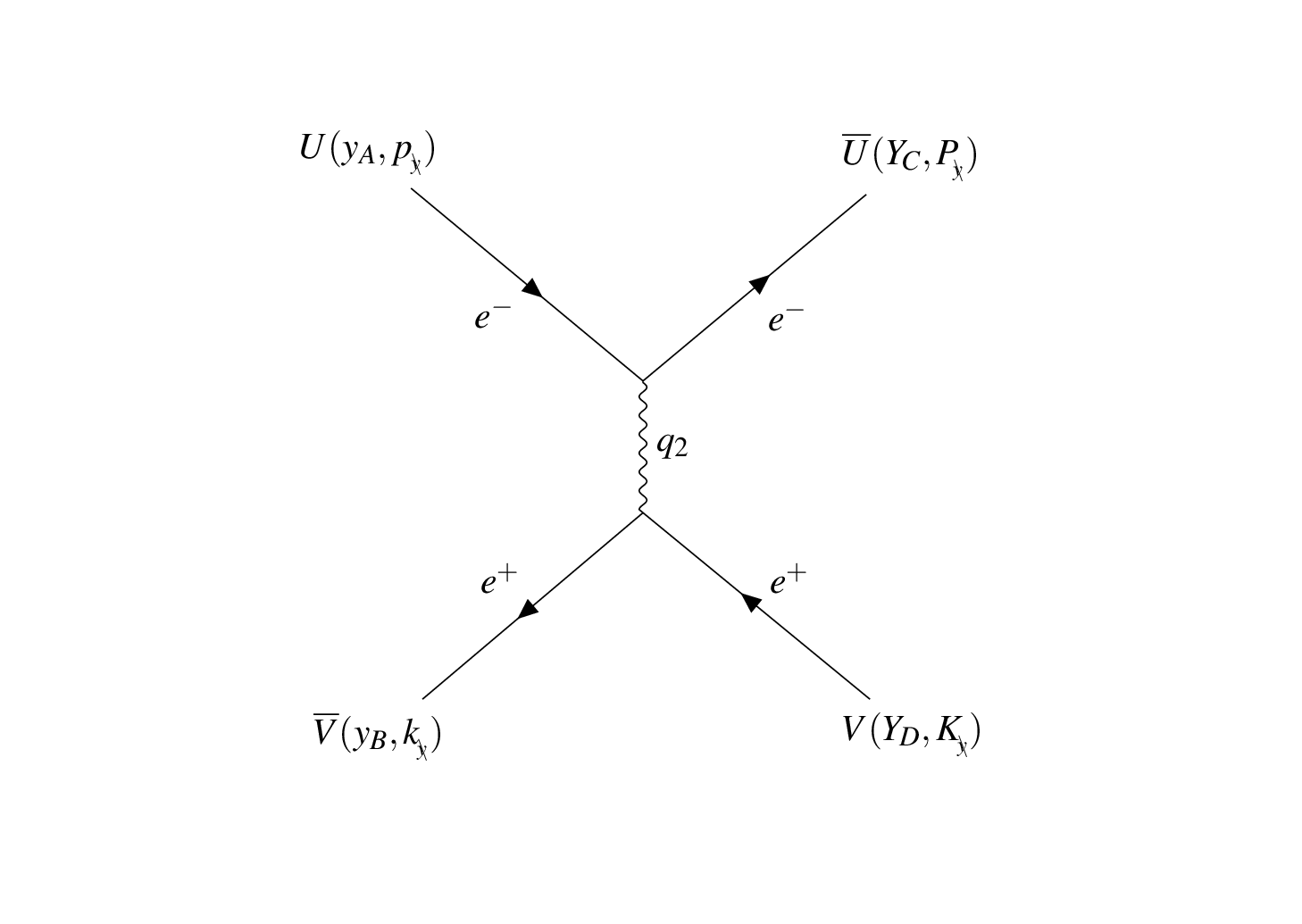}
		\vspace{-1cm}
			\caption{$t$-channel}\label{fd4}
\end{minipage}
\end{figure}

There are possible $s$- and $t$-channel diagrams, which contribute
to the Bhabha scattering, in Figure(s) \ref{s-ch} and \ref{fd4}, 
respectively in the lowest order.
Therefore the matrix element for the Bhabha scattering in the 
lowest order is
\begin{eqnarray}
{\mathfrak{M}}\bhabha = {\mathfrak{M}}_{\rm{s}}\bhabha+ 
{\mathfrak{M}}_{\rm{t}}\bhabha ,
\end{eqnarray}
where the $s$- and $t$-channel 
contributions to the matrix element in a strong magnetic field are
given by
\begin{eqnarray}
{\mathfrak{M}}_{\rm{s}}\bhabha 
 &=& \frac{-e^2}{q_1^2}\big[\overline{V}(y_B,k_{\omit{y}})
\gamma^\mu U(y_A,p_{\omit{y}})\big]\big[\overline{U}(Y_C,P_{\omit{y}})
\gamma_\mu V(Y_D,K_{\omit{y}})\big], \label{bhabha-m-s}\\
\mathfrak{M}_{\rm{t}} (e^- e^+ \rightarrow e^- e^+)
&=& \frac{-e^2}{q_2^2}\big[\overline{U}(Y_C,P_{\omit{y}})
\gamma^\mu U(y_A,p_{\omit{y}})\big]\big[\overline{V}(y_B,k_{\omit{y}})
\gamma_\mu V(Y_D,K_{\omit{y}})\big],\label{bhabha-m-t}
\end{eqnarray}
respectively. Hence the total matrix element squared becomes
\begin{equation}
|\mathfrak{M}|^2=|\mathfrak{M}_{\rm{s}}|^2+|\mathfrak{M}_{\rm{t}}|^2 +
\mathfrak{M}_{\rm{s}}\mathfrak{M}_{\rm{t}}^*+\mathfrak{M}_{\rm{s}}^*
\mathfrak{M}_{\rm{t}}.
\end{equation} 
Similar to the electron-muon scattering in the $s$-channel~
\eqref{matrix-element-square-3}, we can now directly write the 
spin-summed
squared matrix element of the Bhabha scattering for the $s$-channel 
in extreme relativistic limit
\begin{eqnarray}
\overline{\left|\mathfrak{M}_{\rm{s}}\right|^2}{\bhabha}&=&\frac{2e^4}{q_1^4}\Big[(p_\parallel\cdot K_\parallel)(k_\parallel\cdot P_\parallel)+(p_\parallel\cdot P_\parallel)(k_\parallel\cdot K_\parallel)+(p_\parallel\cdot \widetilde K_\parallel)(k_\parallel\cdot \widetilde P_\parallel)\nonumber\\
&&+(p_\parallel\cdot \widetilde P_\parallel)(k_\parallel\cdot \widetilde K_\parallel)+(\widetilde p_\parallel\cdot K_\parallel)(\widetilde k_\parallel\cdot P_\parallel)+(\widetilde p_\parallel\cdot P_\parallel)(\widetilde k_\parallel\cdot K_\parallel)\nonumber\\
&&+(\widetilde p_\parallel\cdot \widetilde K_\parallel)(\widetilde k_\parallel\cdot \widetilde P_\parallel)+(\widetilde p_\parallel\cdot \widetilde P_\parallel)(\widetilde k_\parallel\cdot \widetilde K_\parallel)\Big],
\end{eqnarray}
which in terms of Mandelstam and Magnetic Mandelstam variables can 
be written as
\begin{equation}
\overline{\left|\mathfrak{M}_{\rm{s}}\right|^2}{\bhabha}=\frac{e^4}{s^2}\Big[u^2+t^2+u_pu_k+t_pt_k\Big].\label{bhabha_s}
\end{equation}
The contribution for the Bhabha scattering in $t$-channel can be obtained in 
extreme relativistic limit, in analogy with 
$e$-$\mu$ scattering for $t$-channel in \eqref{e-u-t-ch},
\begin{eqnarray}
\overline{\left|\mathfrak{M}_{\rm{t}}\right|^2}{\bhabha}&=&\frac{2e^4}{q_2^4}\Big[(p_\parallel\cdot k_\parallel)(P_\parallel\cdot K_\parallel)+(p_\parallel\cdot K_\parallel)(P_\parallel\cdot k_\parallel)+(p_\parallel\cdot \widetilde k_\parallel)(P_\parallel\cdot \widetilde K_\parallel)\nonumber\\
&&+(p_\parallel\cdot \widetilde K_\parallel)(P_\parallel\cdot \widetilde k_\parallel)\nonumber+(\widetilde p_\parallel\cdot k_\parallel)(\widetilde P_\parallel\cdot K_\parallel)+(\widetilde p_\parallel\cdot K_\parallel)(\widetilde P_\parallel\cdot k_\parallel)\nonumber\\
&&+(\widetilde p_\parallel\cdot \widetilde k_\parallel)(\widetilde P_\parallel\cdot \widetilde K_\parallel)
+(\widetilde p_\parallel\cdot \widetilde K_\parallel)(\widetilde P_\parallel\cdot \widetilde k_\parallel)\Big],
\end{eqnarray}
which can be expressed in terms of Mandelstam variables as
\begin{equation}
\overline{\left|\mathfrak{M}_{\rm{t}}\right|^2}{\bhabha}=
\frac{e^4}{t^2}\Big[s^2+u^2+s_ps_P+u_pu_P\Big].\label{bhabha_t}
\end{equation}
The interference term is given by
\begin{eqnarray}
\mathfrak{M}_{\rm{s}}\mathfrak{M}_{\rm{t}}^*&=&\frac{e^4}{q_1^2q_2^2}\big[\overline{V}(y_B,k_{\omit{y}})\gamma^\mu U(y_A,p_{\omit{y}})\big]\big[\overline{U}(Y_C,P_{\omit{y}})\gamma_\mu V(Y_D,K_{\omit{y}})\big]\nonumber\\
&&\times \big[\overline{V}(Y_D,K_{\omit{y}})\gamma_\nu V(y_B,k_{\omit{y}})
\big] \big[\overline{U}(y_A,p_{\omit{y}})\gamma^\nu U(Y_C,P_{\omit{y}})
\big].
\end{eqnarray}
However, in the strong magnetic field limit, all the spatial ($y$) 
dependence 
of the Dirac spinors are gone and also the $p_\perp$ 
is zero, so the above interference term is rewritten as
\begin{eqnarray}
	\mathfrak{M}_{\rm{s}}\mathfrak{M}_{\rm{t}}^*&=&\frac{e^4}{q_1^2q_2^2}\big[\overline{V}(k_\parallel)\gamma^\mu U(p_\parallel)\big]\big[\overline{U}(P_\parallel)\gamma_\mu V(K_\parallel)\big]\big[\overline{V}(K_\parallel)\gamma_\nu V(k_\parallel)\big] \big[\overline{U}(p_\parallel)\gamma^\nu U(P_\parallel)\big],
\end{eqnarray}
which becomes, after summing over the spin states
\begin{eqnarray}
\overline{\mathfrak{M}_s\mathfrak{M}_t^*}
	=\frac{e^4}{q_1^2q_2^2}Tr\Big[P_V(k_\parallel)\gamma^\mu P_U 
(p_\parallel)\gamma^\nu P_U(P_\parallel)\gamma_\mu P_V(K_\parallel)
\gamma_\nu\Big]. 
\end{eqnarray}
Using the property of $\gamma$-matrices~
$\gamma^\mu\slashed{a}\slashed{b}\slashed{c}\gamma_\mu=-2\slashed{c}
\slashed{b}\slashed{a}$, the above interference term can be further
simplified in terms of Mandelstam and the magnetic Mandelstam 
variables\footnote{Calculated in Appendix B} and finally 
we find that it vanishes, 
\begin{eqnarray}
\overline{\mathfrak{M}_{\rm s}\mathfrak{M}_{\rm t}^*} &=&0\label{M1M2}.
\end{eqnarray}
Finally using the Mandelstam and magnetic Mandelstam variables, the matrix 
element squared for the Bhabha scattering is obtained by the $s$- \eqref{bhabha_s} 
and $t$-channel \eqref{bhabha_t} contributions only in a strong magnetic field
\begin{eqnarray}
{\overline{\left|\mathfrak{M}\right|^2}}_{\rm B \neq 0} \bhabha&=&
\overline{\left|\mathfrak{M}_{\rm s}\right|^2} +
\overline{\left|\mathfrak{M}_{\rm t}\right|^2} \nonumber\\
&=&\frac{e^4}{s^2}\Big[u^2+t^2+u_pu_k+t_pt_k\Big]+\frac{e^4}{t^2}\Big[s^2+u^2+s_ps_P+u_pu_P\Big].\label{Bhabha_M}
\end{eqnarray}

The above crucial observation in a strong magnetic field can be 
understood
as follows: In vacuum, Bhabha scattering in $s$ channel gives the 
forward peak 
whereas in $t$-channel it gives the backward peak. In the presence of
strong magnetic field, the dynamics of the electron is restricted 
to one 
dimension so the interference of two peaks in vacuum may not be
feasible in the presence of strong magnetic field.

However, in vacuum, the interference term does not vanish, thus the above 
matrix element squared for Bhabha scattering in the 
absence of strong magnetic field is given by~\cite{halzen_martin}
\begin{eqnarray}
{\overline{\left|\mathfrak{M}\right|^2}}_{\rm B =0} \bhabha=
\frac{2e^4}{s^2}\Big[u^2+t^2\Big]+\frac{4e^4u^2}{ts}+
\frac{2e^4}{t^2}\Big[s^2+u^2\Big].\label{bhabha-vacuum-m}
\end{eqnarray}

\subsection{Matrix Element for M\o ller Scattering }
The Feynman diagrams for the M\o ller scattering ($e^-e^- \rightarrow 
e^-e^-$) are shown below in Figure (s) \ref{moller_fd_1} and \ref{moller_fd_2}, 
\begin{figure}[h]
	\begin{minipage}{0.5\textwidth}
		\includegraphics[height=9cm,width=7cm,angle=-90]{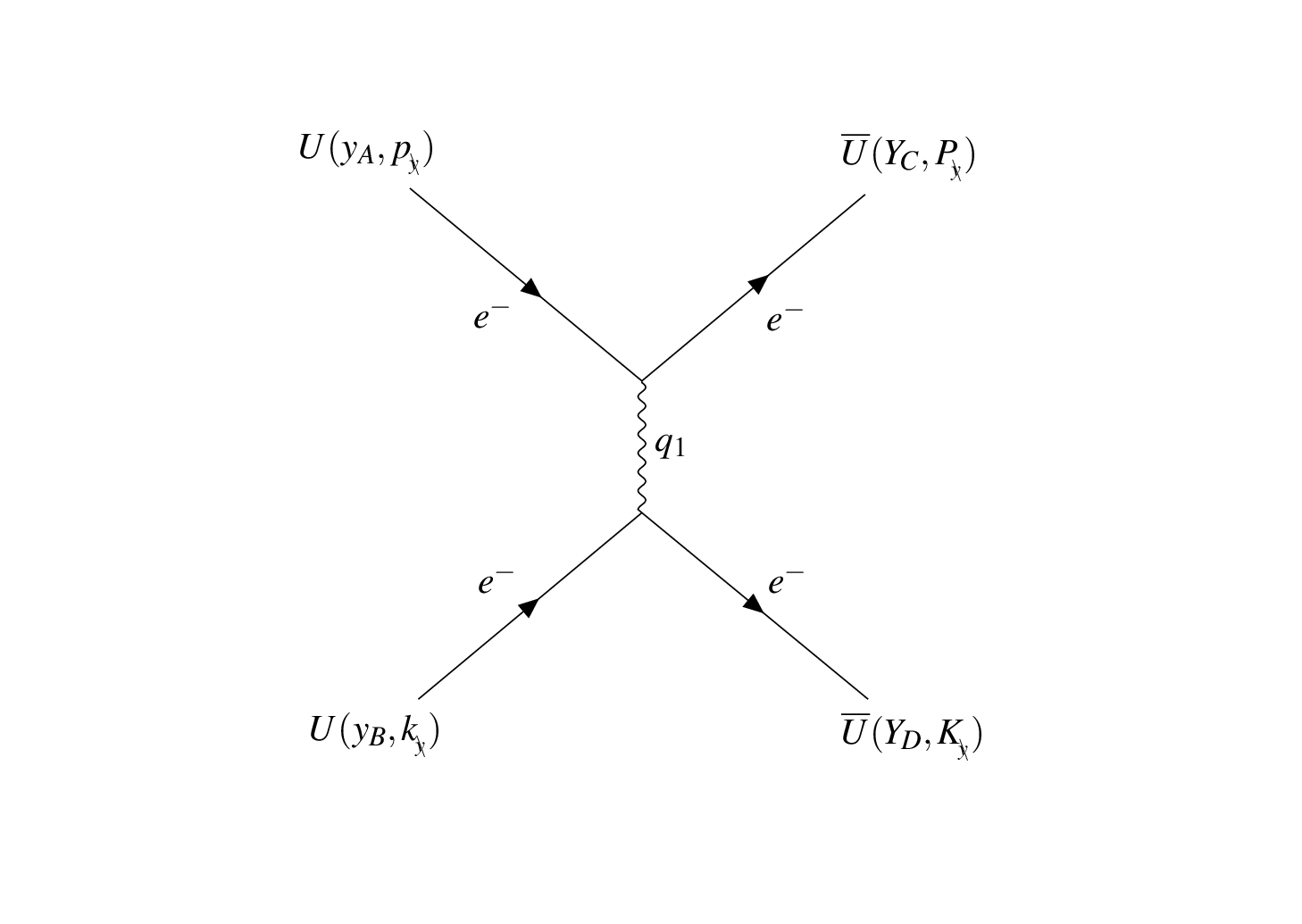}
		\vspace{-1cm}
		\caption{$t$-channel}\label{moller_fd_1}
	\end{minipage}
	\begin{minipage}{0.5\textwidth}
		\includegraphics[height=9cm,width=7cm,angle=-90]{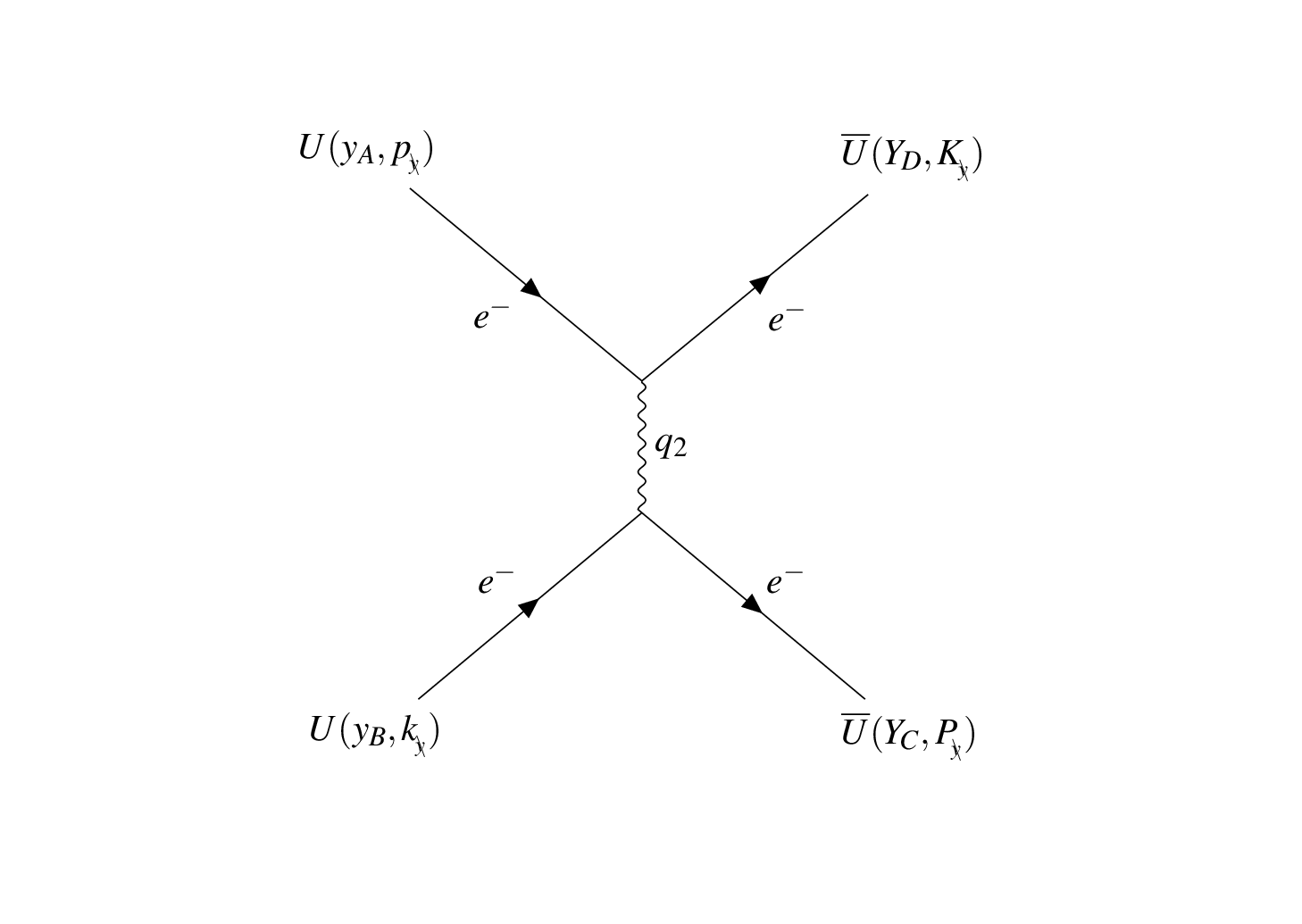}
		\vspace{-1cm}
		\caption{$u$-channel}\label{moller_fd_2}
	\end{minipage}
\end{figure}
which are related to the Bhabha scattering ($e^-e^+ 
\rightarrow e^-e^+$) by the crossing symmetry, {\em namely}
by simply crossing the incoming positron to outgoing positron
\begin{eqnarray}
&&e^-(p)+e^+(k) \rightarrow e^-(P)+e^+(K),\label{bhabha-eq}\\
&&e^-(p)+e^-(-K) \rightarrow e^-(P)+e^-(-k).\label{moller-eq}
\end{eqnarray}
As we know already, in strong magnetic field, the interchange of momenta due to the crossing
symmetry is effectively translated in terms of their longitudinal 
component only. Thus the momentum exchange between $k_\parallel$ and $K_\parallel$ in 
\eqref{bhabha-eq}-\eqref{moller-eq} helps to obtain the matrix element for the
M\o ller scattering in u and t-channel by identifying the same for the Bhabha scattering in s- and
t-channels \eqref{bhabha-m-s}-\eqref{bhabha-m-t}, respectively.

The above interchange of momenta can be equivalently expressed in terms of the Mandelstam
and magnetic Mandelstam variables as
\begin{eqnarray}
s (=(p_\parallel+k_\parallel)^2) &\longrightarrow& u (=(p_\parallel-
K_\parallel)^2)\label{moller-cross-1}\\
u (=(p_\parallel-K_\parallel)^2) &\longrightarrow& s(=(p_\parallel+
k_\parallel)^2)\\
t_k (=(\widetilde{k}_\parallel-K_\parallel)^2) &\longrightarrow& t_K (= 
(k_\parallel-\widetilde{K}_\parallel)^2)\\
s_k(=(p_\parallel+\widetilde k_\parallel)^2) &\longrightarrow& u_K(= 
(p_\parallel-\widetilde K_\parallel)^2).\label{moller-cross-2}
\end{eqnarray}

Therefore the above crossing symmetry \eqref{moller-cross-1}-\eqref{moller-cross-2} helps 
us to obtain the squared 
matrix element for the M\o ller scattering from the Bhabha 
scattering~\eqref{Bhabha_M}, {\em namely}
\begin{eqnarray}
{\overline{\left|\mathfrak{M}\right|^2}}_{\rm B \neq 0} {\moller} &=&
{\overline{\left|\mathfrak{M}_{\rm u}\right|^2}}_{\rm B \neq 0} {\moller}+
{\overline{\left|\mathfrak{M}_{\rm t} \right|^2}}_{\rm B \neq 0}{\moller}, 
\end{eqnarray}
where the $u$ and $t$-channel matrix element squared are given by
\begin{eqnarray}
{\overline{\left|\mathfrak{M}_{\rm u}\right|^2}}_{\rm B \neq 0} \moller &=&
\frac{e^4}{u^2}\Big[s^2+t^2+s_ps_K+t_pt_K\Big],\label{moller-u} \\
{\overline{\left|\mathfrak{M}_{\rm t} \right|^2}}_{\rm B \neq 0} {\moller} &=& 
\frac{e^4}{t^2}\Big[u^2+s^2+u_pu_P+s_ps_P\Big],\label{Moller_M}
\end{eqnarray}
respectively. However, the above matrix element in vacuum can also
be calculated as~\cite{halzen_martin}
\begin{eqnarray}
{\overline{\left|\mathfrak{M}\right|^2}}_{\rm B =0} \moller=
\frac{2e^4}{u^2}\Big[s^2+t^2\Big]+\frac{4e^4s^2}{tu}+
\frac{2e^4}{t^2}\Big[s^2+u^2\Big].\label{moler-vacuum-m}
\end{eqnarray}

\section{Crosssection}
\begin{figure}[h]
	\begin{center}
		\includegraphics[height=12cm,width=8cm,angle=-90]{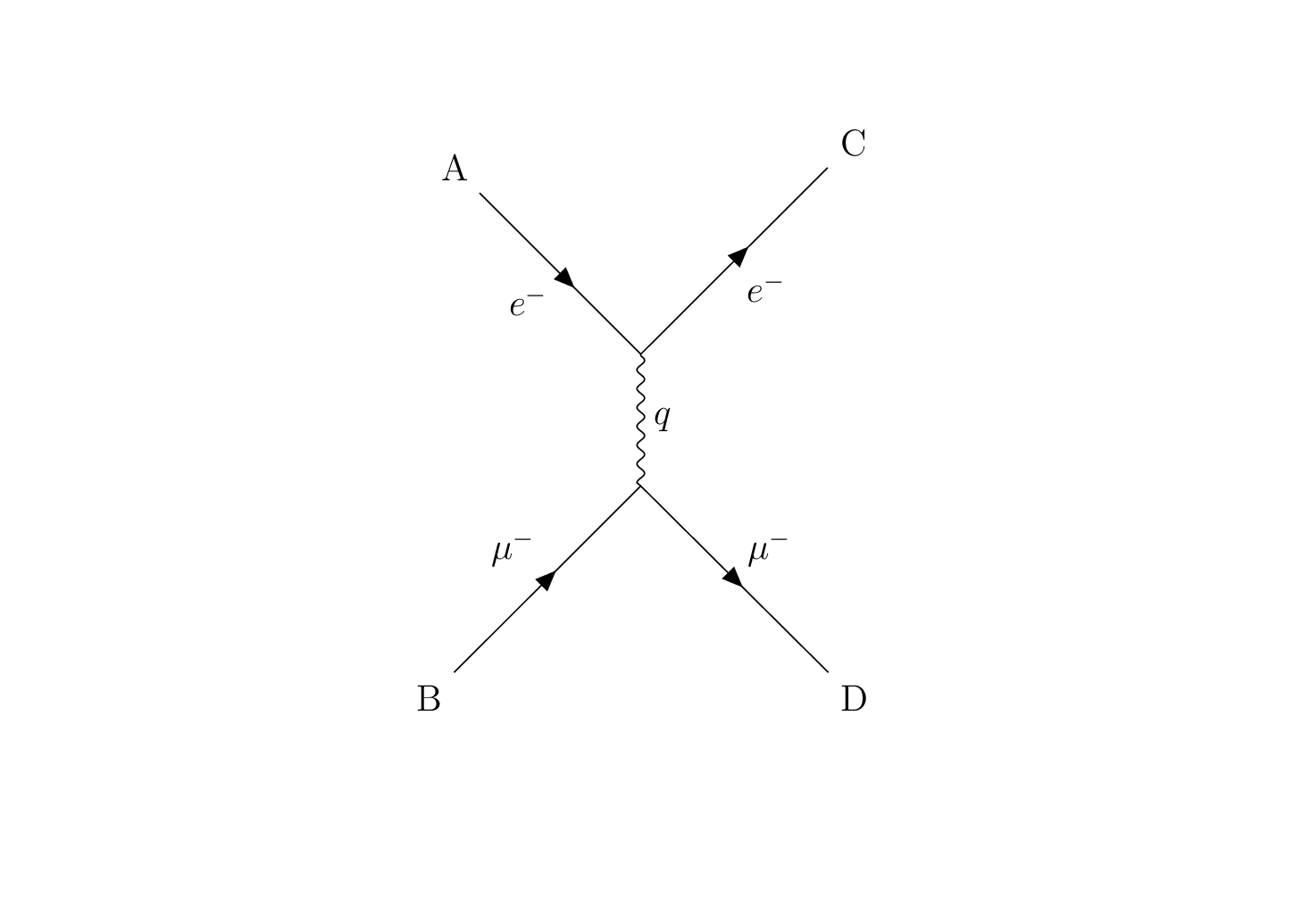}
		\vspace{-1cm}
		\caption{Feynman diagram}\label{lifs}
	\end{center}
\end{figure}

Let us illustrate the usual procedure to compute the
crosssection from the transition amplitude for the
above mentioned processes in the presence of strong magnetic field.
For that we choose a generic $e^-\mu^-
\rightarrow e^-\mu^-$ process in Figure \ref{lifs}, for which we will calculate the 
transition amplitude.

The solution of the Dirac equation for the $e^-$ in an external 
magnetic field in the $z$-direction is given 
by~\cite{Furry,Bhattacharya:2007vz}
\begin{equation*}
	\Psi(X,p)=U(y,n,p_{\omit{y}}) e^{-ip_{\omit{y}}\cdot X},
\end{equation*}
where $U$ is the $e^-$ spinor in the presence of an external magnetic field, $n$ 
labels
the Landau levels, $X^\mu=(t,x,y,z)$ and $p_{\omit{y}}^\mu=(E,p_x,0,p_z)$,
where the $y$-component of momentum is missing.  The transition 
matrix 
element for the above process is thus given by
\begin{eqnarray}
	T_{fi}&=&\int (\overline{U}_D ie\gamma^\mu U_B) \left(\frac{-i}{q^2}\right)(\overline{U}_C ie\gamma_\mu U_A ) e^{i(p^D_{\omit{y}}+p^C_{\omit{y}}-p^A_{\omit{y}}-p^B_{\omit{y}})\cdot X}d^4X\\
	&\equiv &\int- i \hspace{1mm}\mathfrak{M}\hspace{1mm} 
e^{i(p^D_{\omit{y}}+p^C_{\omit{y}}-p^A_{\omit{y}}-p^B_{\omit{y}})\cdot X}
d^4X.
\end{eqnarray}

Therefore the matrix element is given by
\begin{equation}
 -i \mathfrak{M}= (\overline{U}_D ie\gamma^\mu U_B) \left(\frac{-i}{q^2}\right)
(\overline{U}_C ie\gamma_\mu U_A ),
\end{equation}
where the main problem arises that $
\mathfrak{M}$ now becomes a function of $y$ in the presence of
an external magnetic field, hence we can not take it outside the 
integration. We circumvent this issue by taking the strong magnetic 
field limit, where all the $y$ dependence 
in $\mathfrak{M}$ is gone. The transition matrix element, 
in the strong magnetic field becomes 
\begin{eqnarray}
	T_{fi} =- i \hspace{1mm}\mathfrak{M} \hspace{1mm}(2\pi)^4 
\delta^4(p^D_{\omit{y}}+p^C_{\omit{y}}-p^A_{\omit{y}}-p^B_{\omit{y}}),
\end{eqnarray}
and the transition rate per unit volume is given by
\begin{eqnarray}
W_{fi}&=&\frac{|T_{fi}|^2}{TV}\nonumber\\
&=&(2\pi)^4|\mathfrak{M}|^2 \delta^4(p^D_{\omit{y}}+p^C_{\omit{y}}-
p^A_{\omit{y}}-p^B_{\omit{y}}).
\end{eqnarray}

Therefore using the definition of crosssection 
\begin{equation*}
\mbox{crosssection}=\frac{W_{fi}}{\mbox{(initial flux)}}\mbox{(number of 
final states)},
\end{equation*}
the crossection is given by
\begin{eqnarray}
d\sigma=\frac{(2\pi)^4|\mathfrak{M}|^2 \delta^4(p^D_{\omit{y}}+p^C_{\omit{y}}-p^A_{\omit{y}}-p^B_{\omit{y}})}{F}d^4p^Cd^4p^D.
\end{eqnarray} 
All the particles A,B,C and D are real particles so they 
must satisfy the on-shell mass condition, which in the  
strong magnetic field becomes $p_\parallel^2=m^2$ 
\cite{Gusynin:1998nh}
with $p_\parallel^\mu=(E,0,0,p_z)$. This gives
the crosssection 
\begin{eqnarray}
	d\sigma=\frac{(2\pi)^4|\mathfrak{M}|^2 \delta^4(p^D_{\omit{y}}+ 
p^C_{\omit{y}}-p^A_{\omit{y}}-p^B_{\omit{y}})}{F}
\frac{d^3\vec{p^C}}{(2\pi)^32E_C}\frac{d^3\vec{p^D}}{(2\pi)^32E_D}. \label{liff}
\end{eqnarray} 

To calculate the unpolarized crosssection, we need to average over the 
quantum states of incoming particles and sum over the final states, 
therefore we replace the above matrix element squared 
\begin{equation}
|\mathfrak{M}|^2 \rightarrow \frac{1}{(2s_A+1)(2s_B+1)}\sum_{all ~states}|\mathfrak{M}|^2,
\end{equation} 
where $s_A$ and $s_B$ are the spin of incoming particles.
We have already summed over the particle states in the matrix 
element squared,
denoted as $\overline {\left|\mathfrak{M}\right|^2}$, so we just 
need to divide 
$\overline {\left|\mathfrak{M}\right|^2}$ by the degeneracy factor, 4, to get the 
desired crosssection.

Now we have all the ingredients to compute the crosssection for the 
aforesaid 
processes with the corresponding matrix element squared.

\subsection{Electron-Muon scattering in $s$ channel: 
Annihilation Process ($e^-e^+\rightarrow \mu^-\mu^+$)}
With the help of~\eqref{liff} the crosssection for the 
electron-muon scattering in
$s$-channel ($e^-e^+\rightarrow \mu^-\mu^+$)
in Figure \ref{s_e_mu} with the matrix element~\eqref{matrix-element-square-4} is
\begin{eqnarray}
d\sigma^{\rm{s}}_{\rm B \neq 0 } \ems
&=&\frac{(2\pi)^4}{F} \overline{\left|\mathfrak{M}_{\rm s}\right|^2} \ems \delta^4(K_{\omit{y}}+P_{\omit{y}}-k_{\omit{y}}-p_{\omit{y}})\frac{d^3\vec{P}}{(2\pi)^32E_P}\frac{d^3\vec{K}}{(2\pi)^32E_K}\nonumber\\
	&=&\frac{ \overline{\left|\mathfrak{M}_{\rm s}\right|^2}\ems}{(2\pi)^24F} \delta(E_P+E_K-E_p-E_k)\delta^3(\vec{K}_{\omit{y}}+\vec{P}_{\omit{y}}-\vec{k}_{\omit{y}}-\vec{p}_{\omit{y}})\nonumber\\
	&&\hspace{10cm}\times\frac{d^3\vec{P}}{E_P}\frac{d^3\vec{K}}{E_K}.\nonumber
\end{eqnarray}

In the center-of-mass (cm) frame, $\vec{p}+\vec{k}=0$, this also implies, $\vec{p}_{\omit{y}}+\vec{k}_{\omit{y}}=0$. 
Thus then crosssection becomes
\begin{eqnarray}
	d\sigma^{\rm{s}}_{\rm B \neq 0} \ems&=&\frac{ \overline{\left|\mathfrak{M}_{\rm s}\right|^2}}{(2\pi)^24F} \delta(E_P+E_K-E_p-E_k)\delta^3(\vec{K}_{\omit{y}}+\vec{P}_{\omit{y}})\frac{d^3\vec{P}}{E_P}\frac{d^3\vec{K}}{E_K}\\
	&=&\frac{ \overline{\left|\mathfrak{M}_{\rm s}\right|^2}_{\vec{K}_{\omit{y}}=-\vec{P}_{\omit{y}}}}{(2\pi)^24F} \delta(E_P+E_K-E_p-E_k)\frac{d^3\vec{P}}{E_KE_P}.
\end{eqnarray}

In the presence of magnetic field, the momentum integration gets
factorized into parallel and perpendicular components 
with respect to the direction of magnetic field ($z$-direction),
where the integral over $d^2P_\perp$ in strong 
magnetic field limit becomes ($\int_0^{|eB|}d^2P_\perp
=\pi |eB|$). Thus the total crosssection is obtained 
by integrating over the parallel component of momentum($P_z$) 
\begin{eqnarray}
\sigma^{\rm{s}}_{\rm B \neq 0} \ems &=&\int_{-\infty}^\infty \frac{\pi|eB| \overline{\left|\mathfrak{M}_{\rm s}\right|^2}_{K_z=-P_z}}{(2\pi)^24F} \delta(E_P+E_K-E_p-E_k)\frac{dP_z}{E_KE_P}.\label{d-sig} 
\end{eqnarray}

\begin{figure}[h]
	\begin{center}
	\includegraphics[height=12cm,width=8cm,angle=-90]{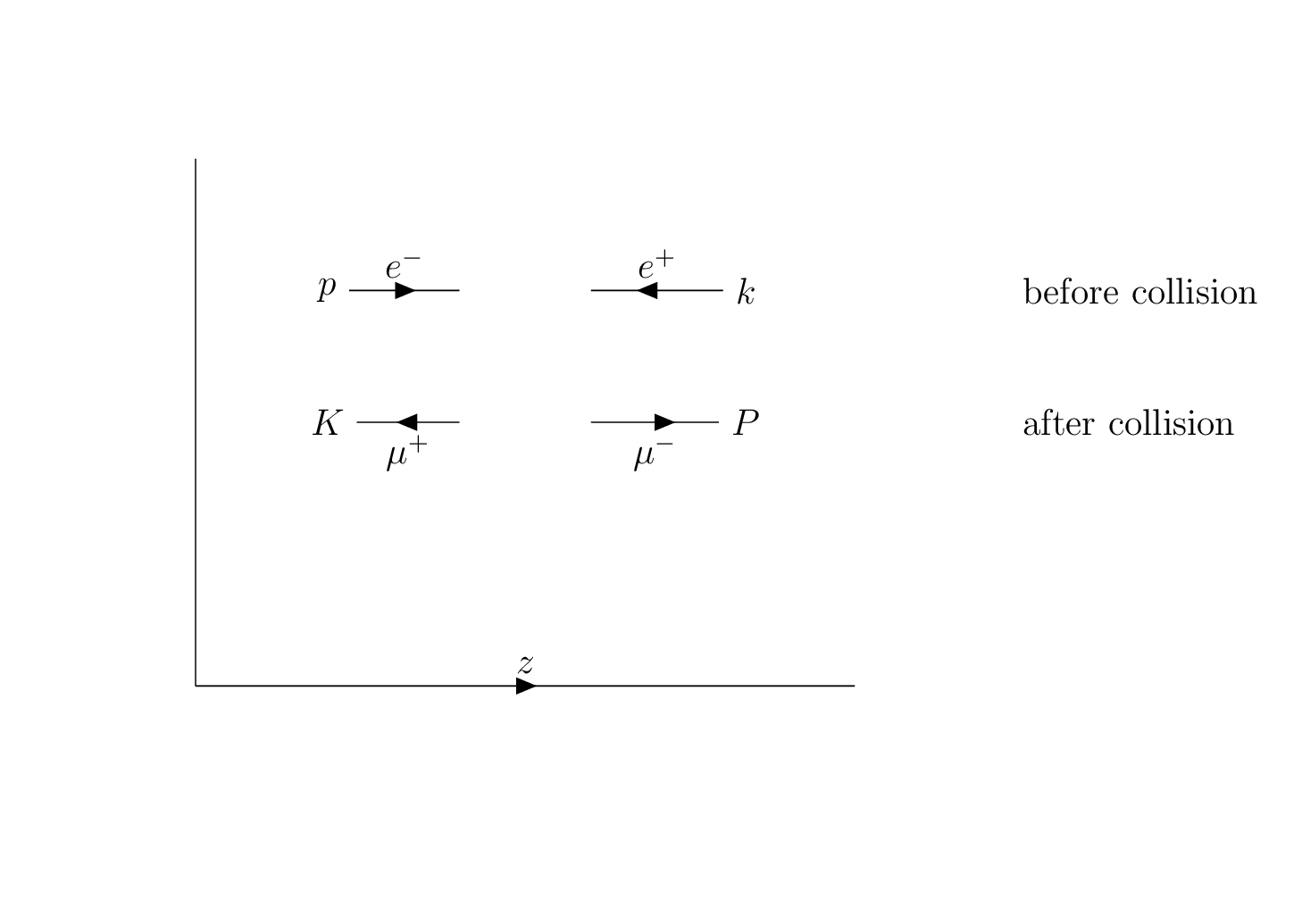}
	\vspace{-1cm}
	\caption{Reaction in center-of-mass frame}\label{ricf}	
	\end{center}
\end{figure}

In a strong magnetic field limit ($eB>>m^2; n=0$), the perpendicular
component ($\perp$) of the momentum is zero~\cite{Gusynin:1998nh} so 
the particles can only move in $z$ direction. They can either move 
in $+ve$ $ z$ direction or in $-ve$ $ z$ direction.  Accordingly 
the four momentum dot product can be written as
\begin{equation}
p.k=
\begin{cases}
E_pE_k-|\vec{p}||\vec{k}|,& \mbox{if $\vec{p}$ and $ \vec{k}$ are in 
the same direction} \\
E_pE_k+|\vec{p}||\vec{k}|,& \mbox{if $\vec{p}$ and $ \vec{k}$ are in 
the opposite direction.} 
\end{cases}
\end{equation}
In extreme relativistic limit we can neglect the dot product where the momenta 
are in the same direction compared to them in opposite direction.
The diagram for the process in the center-of-mass frame (for 
$\theta=0$ degree, where 
$\theta$ is the angle between $p$ and $P$) is 
drawn in the Figure \ref{ricf}.
For our case $\theta$ (scattering angle) can have two values $0$ 
and $180$ degree, for 
$\theta=0$ the $t$ variables in eq~\eqref{matrix-element-square-4} are 
negligible as compared to all $u$ variables whereas for $\theta=180$
the $u$ variables are negligible as compared to all $t$ variables.
Thus the squared matrix element eq~\eqref{matrix-element-square-4} 
at high energy for $\theta=0$ degree
gets simplified 
\begin{eqnarray}
{\overline{\left|\mathfrak{M}_{\rm s}\right|^2}}_{\rm B \neq 0} (e^- e^+ 
\rightarrow \mu^- \mu^+) 
&=&\frac{e^4}{4s^2}\Big[u^2+u_pu_k\Big],
\end{eqnarray} 
and for $\theta=180$ degree, it becomes
\begin{eqnarray}
{\overline{\left|\mathfrak{M}_{\rm s}\right|^2}}_{\rm B \neq 0} (e^- e^+ 
\rightarrow \mu^- \mu^+) &=&\frac{e^4}{4s^2}\Big[t^2+t_pt_k\Big].
\end{eqnarray}

Let us denote $|\vec{p}|=|\vec{k}|=p_i$ and 
$|\vec{P}|=|\vec{K}|=p_f$ so
 $E_p$ and $E_k$ become the same (say, $E_i$) whereas $E_P$ and $E_K$ 
become equal (say, $E_f$). As a consequence the square of the matrix 
element for $\theta=0$ degree 
comes out to be same as for the $\theta=180$ degree. Hence 
\begin{equation}
\overline{\left|\mathfrak{M}_{\rm s}\right|^2}{\ems}
=\frac{2e^4}{s^2}E_f^2\left[E_i+p_i \right]^2,\label{ep_in_cm}
\end{equation}
and the flux factor for collinear collision is
\begin{eqnarray}
	F&=&|\vec{v}_p-\vec{v}_k|2E_p2E_k\nonumber\\
	&=&4p_i\sqrt{s},
\end{eqnarray}
where $\sqrt{s}=E_p+E_k$.

Using the expression of the flux factor $F$ and the squared matrix 
element
$\overline{\left|\mathfrak{M}_{\rm s}\right|^2}$, the crosssection
\eqref{d-sig} can be rewritten as
\begin{eqnarray}
\sigma^{\rm{s}}_{\rm B \neq 0} {\ems}=\frac{e^4|eB|}{16\pi} \int_{-\infty}^{\infty}
	\frac{\pi e^4|eB|\left[E_i+p_i \right]^2}{s^2 p_i\sqrt{s}} 
\delta(W-\sqrt{s})\frac{E_f^2}{p_fW}dW,
\end{eqnarray}
where $W=E_P+E_K$.

With the further approximation: $p_f \approx E_f$ 
and $p_i \approx E_i$, the crosssection for the $e$-$\mu$ 
scattering in 
$s$-channel\footnote{Detailed calculation of crosssection for the 
process $e^-e^+\rightarrow \mu^-\mu^+$ is given in Appendix C}, 
{\em i.e.}
for the annihilation process ($e^-e^+\rightarrow \mu^-\mu^+$) in the
lowest order takes the final form as a function of the 
center-of-mass energy
\begin{eqnarray}
\sigma^{\rm{s}}_{\rm B \neq 0} (e^-e^+\rightarrow \mu^-\mu^+)&=&\frac{\pi \alpha^2 |eB|}{s^2}.
\end{eqnarray}
The approximations, $p_f \approx E_f$ 
and $p_i \approx E_i$ only hold good at extremely high energies, 
where the masses (order of MeV) can be neglected. 

For the sake of comparison, the crosssection for the same 
in vacuum in the lowest order is~\cite{halzen_martin}
\begin{equation}
\sigma^{\rm{s}}_{\rm B=0} (e^-e^+\rightarrow \mu^-\mu^+)=
\frac{4\pi \alpha^2}{3s}. 
\end{equation}
One thus immediately infer that in the 
presence of strong magnetic field, the crosssection 
is inversely proportional to the fourth power 
of the center-of-mass energy while in the absence of magnetic 
field, $\sigma_{\rm B =0}$ 
is inversely proportional to the square of the center-of-mass energy.

\begin{figure}[h]
	\begin{center}
	\includegraphics[height=16cm,width=10cm,angle=-90]{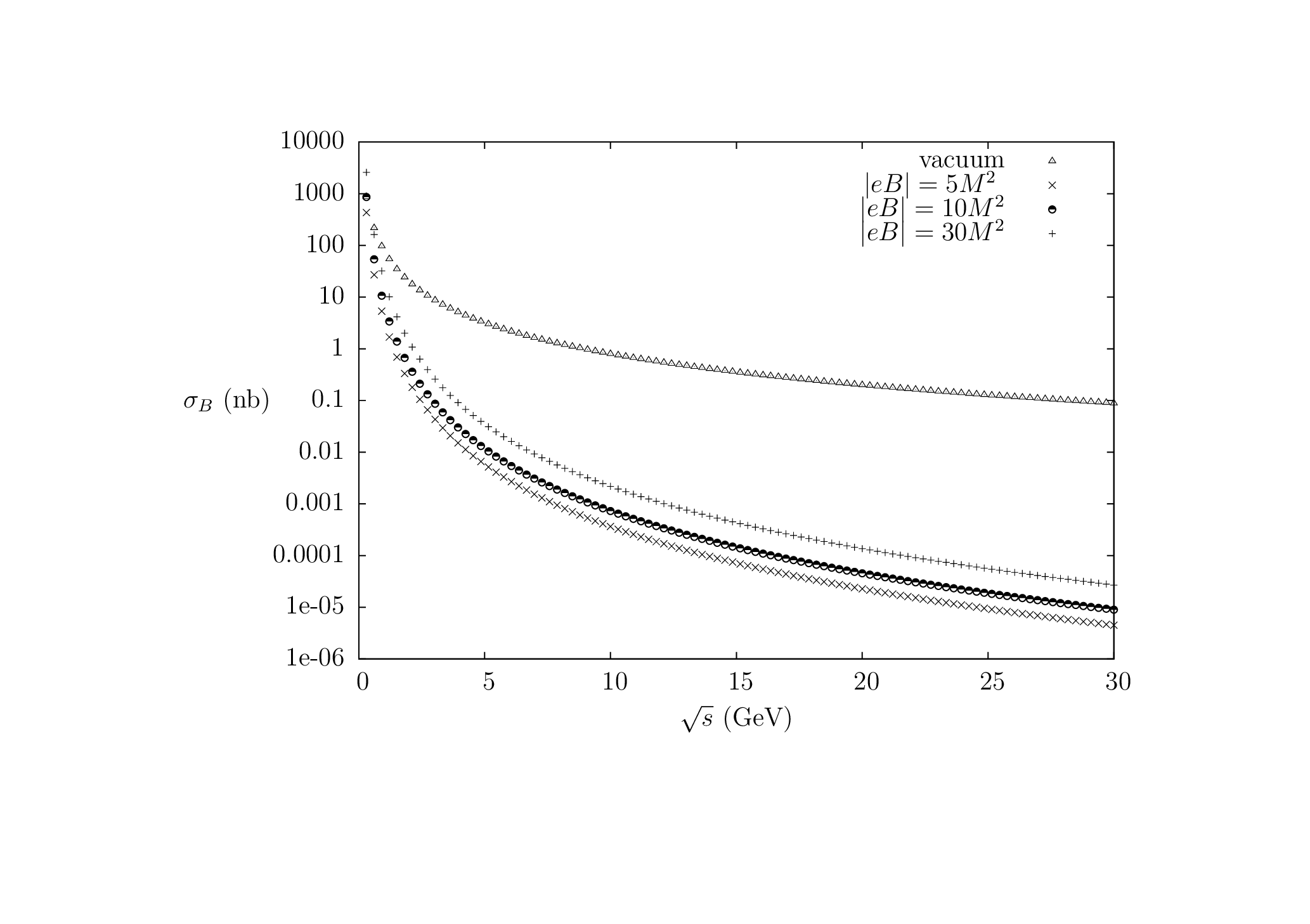}
	\vspace{-2cm}
	\caption{$\sigma_B$ vs center-of-mass Energy at different strength of magnetic field.}\label{sig-plot}
	\end{center}
\end{figure}

To see the effect of strong magnetic field on the annihilation process, we
have plotted a variation of the crosssection with the 
center-of-mass energy in the 
presence and absence of magnetic field in Fig-\ref{sig-plot}, where $M$ is 
the mass of muon.
From the above graph, it is evident that the strong magnetic field 
	suppresses the scattering, which can be understood by the
fact that the availability of phase space in the presence of 
magnetic field 
is reduced drastically. Moreover we have also found that for a
fixed center-of-mass energy, as we increase the magnetic field, 
$\sigma$ increases 
	linearly.

\subsection{Electron-Muon scattering in $t$ channel: $e^-\mu^-\rightarrow e^-
\mu^-$ process}
The crosssection for the electron-muon scattering in $t$ 
channel, {\em i.e.} for the $e^-\mu^-\rightarrow e^-\mu^-$ process 
diverges at $t=0$, since the matrix element in $t$-channel diagram has 
a pole at 
$t=0$. Let us now examine below how the pole $t=0$ translates into the
momentum variable in the final state. The Mandelstam variables, $t$ and $u$ 
for the Figure~\ref{t_e_mu} are defined as
\begin{eqnarray}
t^2&=&4|\vec{p}|^2|\vec{P}|^2(\cos \theta -1)^2=4|\vec{p}|^2|\vec{P}|^2\left(\frac{P_z}{|\vec{P}|}-1\right)^2=4p_i^2p_f^2\left(\frac{P_z}{p_f}-1\right)^2,\label{tt}\\
u^2&=&4|\vec{p}|^2|\vec{K}|^2(\cos \theta +1)^2=4|\vec{p}|^2|\vec{K}|^2
\left(\frac{P_z}{|\vec{P}|}+1\right)^2=4p_i^2p_f^2\left(\frac{P_z}{p_f}
+1\right)^2\label{uu}~,
\end{eqnarray}
where we denote the momenta $|\vec{p}|$ and $|\vec{k}|$ by $p_i$ and 
the momenta $|\vec{P}|$, $|\vec{K}|$ by $p_f$.

As mentioned earlier, in the strong magnetic field limit, 
electrons occupy only the lowest Landau levels ($n=0$), so the lower
limit of the transverse momentum becomes vanishing small, i.e. $P_\perp 
\sim 0$. Therefore, the momentum, $p_f$ (=$|\bf P|=\sqrt{P_\perp^2+P_z^2}$) simply 
becomes $P_z$, hence the pole $t=0$ appears. 
Therefore, using the matrix element for $e$-$\mu$ scattering for $t$-channel
\eqref{matrix-element-squared-t} as well a lower cut-off, $\epsilon_B$ to 
the transverse 
momentum, $P_\perp$ the crosssection \eqref{liff} looks as
\begin{eqnarray}
	\sigma^{\rm{t}}_{\rm B \neq 0}(e^-\mu^-\rightarrow e^-\mu^-)&=&
\int_{\epsilon_B}^{\sqrt{eB}} d P_\perp  \int_{-\infty}^\infty dP_z
	\frac{e^4}{4t^2}\Big[s^2+u^2+s_ks_K+u_Ku_k\Big]
\frac{\delta(W-\sqrt{s})}{(2\pi)4F}\frac{P_\perp}{E_f^2},\qquad
\end{eqnarray}
with the notations: $W=E_P+E_K$, $\sqrt{s}=E_p+E_k$, $E_p=E_k=E_i$ and 
$E_K=E_P=E_f$. Thus after integrating the momentum 
integrations over $dP_\perp$ and $dP_z$, the crosssection
for the $e^-\mu^-\rightarrow e^-\mu^-$ process in the lowest order is
\footnote{Calculated in Appendix D}
\begin{eqnarray}
	\sigma^{\rm{t}}_{\rm B \neq 0}(e^-\mu^-\rightarrow e^-\mu^-)&=&
\frac{\pi \alpha^2}{2s^2}|eB|-\frac{2\pi \alpha^2}{|eB|}+2\pi \alpha^2
\lim_{\epsilon_B \to 0}\left[\frac{1}{\epsilon_B^2}\right]\label{t-sig},
\end{eqnarray}
which implies a divergence in $\epsilon_B \to 0$ limit.
 
For the sake of comparison, we calculate the same crosssection 
in the extreme relativistic limit in the absence of strong magnetic 
field using matrix element from~\eqref{t-vacuum} 
for the $e$-$\mu$ scattering in the $t$-channel
\begin{equation}
\sigma^{\rm{t}}_{\rm B =0} (e^-\mu^-\rightarrow e^-\mu^-)=
-\frac{4\pi \alpha^2}{s}+\frac{4\pi \alpha^2}{s}\lim_{\epsilon_V \to 0}
\left[\frac{2}{\epsilon_V}+\ln{\Big(\frac{\epsilon_V}{2}\Big)}\right].
\end{equation}
This also shows a divergence in the lower limit of $\epsilon_V$, {\em i.e.} 
$\epsilon_V \to 0$, which, in turn, arises due to the  
lower limit of scattering angle (between ${\bf p}$ and ${\bf P}$) as 
\begin{equation}
\epsilon_V=1-\cos \theta_0. 
\end{equation}
However, the above divergences in the presence and the absence of strong 
magnetic field are related to each other and we can geometrically 
derive\footnote{Calculated at the end of Appendix D} an equation 
using the fact that $P_\perp=P_z \tan\theta$, 
which is 
\begin{equation}
\frac{1}{\epsilon_V}=\frac{3}{2}+\frac{s}{2\epsilon_B^2},
\label{magnetic-vacuum}
\end{equation}
Hence, the above relation helps us to compare the crosssection in vacuum
with the crosssection in the presence of magnetic field  
\begin{equation}
\sigma^{\rm{t}}_{\rm B =0} (e^-\mu^-\rightarrow e^-\mu^-)
=\frac{8\pi \alpha^2}{s}+\frac{4\pi \alpha^2}{s}\lim_{\epsilon_V \to 0}
\left[\ln{\Big(\frac{\epsilon_V}{2}\Big)}\right]+4\pi\alpha^2\lim_{\epsilon_B \to 0}\left[\frac{1}{\epsilon_B^2}\right].
\end{equation}

One thus finds that the logarithmic divergence in vacuum disappears due to 
the presence of external magnetic field and apart from that in 
magnetic field there is an another finite term which is independent 
of $s$ but decreases with the increasing magnetic field. 

\subsection{Bhabha Scattering: $e^-e^+\rightarrow e^-e^+$}
The crosssection for the Bhabha Scattering, {\em i.e.} for 
the $e^-e^+\rightarrow e^-e^+$ processes in the lowest-order can be obtained
from the definition \eqref{liff},  with the matrix element \eqref{Bhabha_M}.
In the presence of strong magnetic field, $\sigma (e^-e^+\rightarrow 
e^-e^+$) can be decomposed into $s$- and $t$-channel
contribution due to {\em the vanishing interference term }
\begin{eqnarray}
\sigma_{\rm B \neq 0} (e^-e^+\rightarrow e^- e^+) = \sigma^s_{\rm B \neq 0} (e^-e^+\rightarrow 
e^- e^+) + \sigma^t_{\rm B \neq 0} (e^-e^+\rightarrow e^- e^+),\label{Bhabha_sig_tot}
\end{eqnarray}
where the $s$- and $t$-channel  contribution 
are given by
\begin{eqnarray}
\sigma_{\rm B \neq 0}^{s}\bhabha&=&\frac{e^4}{16\pi s^2}|eB|,\\
\sigma_{\rm B \neq 0}^{t}\bhabha&=&\frac{e^4}{32\pi s^2}|eB|+\frac{e^4}{8\pi} 
\lim_{\epsilon_B \to 0}\left[\frac{1}{\epsilon_B^2}-\frac{1}{|eB|}
\right],
\end{eqnarray}
respectively. Therefore the total crosssection 
\eqref{Bhabha_sig_tot} for the Bhabha 
scattering yields
\begin{eqnarray}
\sigma_{\rm B \neq 0} (e^-e^+\rightarrow e^-e^+)
=\frac{3 \pi \alpha^2}{2 s^2}|eB|-\frac{2\pi \alpha^2}{|eB|}+2 \pi \alpha^2 \lim_{\epsilon_B \to 0}\left[\frac{1}{\epsilon_B^2}\right].\label{bhabha-cross}
\end{eqnarray}
To isolate the effect of strong magnetic field, we have also 
calculated
the same at the lowest-order in vacuum only using the matrix 
element from~\eqref{bhabha-vacuum-m}, where the interference term 
is 
nonzero unlike the former case and is given by
\begin{eqnarray}
\sigma_{\rm B =0} (e^-e^+\rightarrow e^-e^+)&=&\sigma^{\rm s}_{\rm B =0}+\sigma^{\rm interference}_{\rm B =0}+\sigma^{\rm t}_{\rm B =0}\nonumber\\
&=&\left[\frac{4\pi \alpha^2}{3s}\right]-\left[\frac{4\pi \alpha^2}{s}\lim_{\epsilon_V \to 0}\ln{\frac{\epsilon_V}{2}}\right]+\left[\frac{4\pi \alpha^2}{s}\lim_{\epsilon_V \to 0}\left(\frac{2}{\epsilon_V}-1+\ln{\frac{\epsilon_V}{2}}\right)\right]\nonumber\\
&=&-\frac{8\pi \alpha^2}{3s}+\frac{8\pi \alpha^2}{s}\lim_{\epsilon_V \to 0}\left[\frac{1}{\epsilon_V}\right].
\end{eqnarray}

Again using the relation \eqref{magnetic-vacuum}, we can write the above 
crosssection in terms of the parameter ($\epsilon_B$) in the presence of 
strong magnetic field, which causes the divergence 
\begin{eqnarray}
\sigma_{\rm B =0} (e^-e^+\rightarrow e^-e^+)&=&\frac{4\pi \alpha^2}{3s}+
4\pi\alpha^2\lim_{\epsilon_B \to 0}\left[\frac{1}{\epsilon_B^2}\right].
\label{bhabha-vacuum}
\end{eqnarray}
If we leave the divergent part and compare only with the finite part 
then we find that apart from a constant magnetic field dependent 
term, 
the crosssection in vacuum at the lowest order decreases 
with the center-of-mass energy ($\sqrt{s}$) slower than the same in 
the presence of 
strong magnetic field. However, the presence of strong magnetic 
field does not 
alter the degree of divergence.
\subsection{M\o ller Scattering: $e^-e^-\rightarrow e^-e^-$}
The crosssection for the M\o ller scattering at the lowest order 
can be obtained from its matrix element \eqref{Moller_M}, which is
factorizable into $u$ and $t$-channel contributions due to the
vanishing interference term. The crosssection for the $t$-channel 
matrix element in extreme relativistic limit is same as the 
crosssection for electron-muon scattering \eqref{t-sig} in $t$-channel 
$\emt$ process. So
we are left with the crosssection due to the $u$-channel and can be 
calculated from its matrix element \eqref{moller-u} in a strong magnetic field
\begin{eqnarray}
\sigma^{\rm{u}}_{\rm B \neq 0} &=& \int_{-\infty}^\infty dP_z 
\int_0^{\sqrt{eB}} d P_\perp~ \overline{\left|\mathfrak{M}_{\rm u}
\right|^2}\moller \frac{\delta(W-\sqrt{s})}{(2\pi)4F}\frac{P_\perp}{E_f^2},
\end{eqnarray}
with the notations: $W=E_P+E_K$, $\sqrt{s}=E_p+E_k$, $E_p=E_k=E_i$ and 
$E_K=E_P=E_f$. In the above integral, the matrix element has a 
pole at $u=0$,
which could be translated in terms of momentum variable as follows:
In the presence of strong magnetic field, the lower limit of the 
transverse 
momentum ($P_\perp$) becomes vanishingly small. As a result, the momentum, 
$\bf P$  (=$\sqrt{P_\perp^2+P_z^2}$) comes out to be $ \pm P_z$, which give rise
$t=0$ and $u=0$ poles, defined in \eqref{tt} and \eqref{uu}, 
respectively and
can be circumvented by taking a lower cut-off to the lower limit of 
transverse momentum ($P_\perp$) integration.

As an artifact of the above observation, the $u$-channel matrix element 
squared in M\o ller scattering \eqref{moller-u} in the momentum interval,
$P_z \in (-\infty,0]$ is mapped into $t$-channel matrix element \eqref{Moller_M}
in the momentum interval $P_z \in [0,\infty)$ and {\em vice versa}
\begin{eqnarray}
\overline{\left|\mathfrak{M}_{\rm u}\right|^2}_{P_z \in (-\infty,0]}&=&\overline{\left|\mathfrak{M}_{\rm t}\right|^2}_{P_z \in [0,\infty)},\\
\overline{\left|\mathfrak{M}_{\rm u}\right|^2}_{P_z \in [0,\infty)}&=&\overline{\left|\mathfrak{M}_{\rm t}\right|^2}_{P_z \in (-\infty,0]}.
\end{eqnarray} 

As a consequence the crosssections from both channels come out 
to be same except for the fact that in $t$-channel, the 
crosssection peaks
at the forward angle ($\theta=0$) while the $u$-channel peaks
at $\theta=180$. Thus the crosssection for the $\moller$ scattering
in the lowest-order is obtained by doubling the crosssection in 
$t$ channel \eqref{t-sig}
\begin{eqnarray}
\sigma_{\rm B \neq 0} (e^-e^-\rightarrow e^-e^-)&=&\frac{\pi \alpha^2}{s^2}|eB|-\frac{4 \pi \alpha^2}{|eB|}+4 \pi \alpha^2 \lim_{\epsilon_B \to 0}\left[\frac{1}{\epsilon_B^2}\right].
\end{eqnarray}

The crosssection for the M\o ller scattering in vacuum in the 
lowest order using the matrix element from~\eqref{moler-vacuum-m}
can be easily calculated as
\begin{eqnarray}
\sigma_{\rm B=0} (e^-e^-\rightarrow e^-e^-)&=&\sigma_{\rm B =0}^{\rm u}+\sigma_{\rm B =0}^{\rm interference}+\sigma_{\rm B =0}^{\rm t}\nonumber\\
&=&\left[\frac{4\pi \alpha^2}{s}\lim_{\epsilon_V \to 0}\left(\frac{2}{\epsilon_V}-1+\ln{\frac{\epsilon_V}{2}}\right)\right]-\left[\frac{8\pi \alpha^2}{s}\lim_{\epsilon_V \to 0}\ln{\frac{\epsilon_V}{2}}\right]\nonumber\\
&+&\left[\frac{4\pi \alpha^2}{s}\lim_{\epsilon_V \to 0}\left(\frac{2}{\epsilon_V}-1+\ln{\frac{\epsilon_V}{2}}\right)\right]\\
&=&-\frac{8\pi \alpha^2}{s}+\frac{16\pi \alpha^2}{s}\lim_{\epsilon_V \to 0}
\left[\frac{1}{\epsilon_V}\right],
\end{eqnarray}
which can be compared with the result in a strong magnetic field by 
replacing
$\epsilon_V$ in terms of $\epsilon_B$ through the relation \eqref{magnetic-vacuum}
\begin{eqnarray}
\sigma_{\rm B =0} (e^-e^-\rightarrow e^-e^-)&=&\frac{16\pi \alpha^2}{s}+8\pi \alpha^2\lim_{\epsilon_B \to 0}\left[\frac{1}{\epsilon_B^2}\right].
\end{eqnarray}
Like other processes the $s$-dependence in vacuum is being modified due to 
the presence of strong magnetic field whereas the 
diverging component in vacuum ($1\over \epsilon_B^2$) becomes 
halved due to the magnetic field.
\section{Results and Discussions}
We have revisited the lepton-lepton scattering at the lowest order
in an additional presence of strong magnetic 
field. The recent observations at the ultra relativistic 
heavy ion collisions at Relativistic Heavy-ion Collider and Large
Hadron Collider, where
a very strong magnetic field up to ${10}^{18}$ - ${10}^{20}$ Gauss 
is expected to be produced for the noncentral events, motivates
us to revisit the above processes in a strong magnetic field.
In particular, we have calculated the crosssection for 
electron-muon ($e$-$\mu$) scattering in both $s$- and $t$-channel, Bhabha 
scattering, and M\o ller scattering in the presence of a strong 
magnetic field
($|eB|>>m^2$, $m$ is the mass of electron or muon). For that purpose, 
using the Dirac spinor in strong magnetic field, we
have first calculated the square of the matrix element and then 
summed 
over the final spin states. We have found that unlike in vacuum, the 
interference term in Bhabha scattering between $s$- and $t$-channel 
and in M\o ller scattering between $t$- and $u$-channel 
contribution 
in the presence of strong magnetic field vanishes. Secondly we have
illustrated the usual procedure to compute the
crosssection from the transition amplitude for the
above mentioned processes in the presence of strong magnetic field.
We have noticed that the $e$-$\mu$ scattering in $s$-channel,
{\em i.e.} $e^-e^+\rightarrow \mu^-\mu^+$ process gets suppressed
due to the presence of strong magnetic field. More precisely
the crosssection at a fixed magnetic field is inversely 
proportional 
to the fourth power of the center-of-mass energy compared to the 
vacuum alone,
where $\sigma$ is inversely proportional to the square of the 
center-of-mass energy.
However, for a fixed center-of-mass energy, the crosssection 
increases with
the magnetic field.

On the other hand the $e$-$\mu$ scattering in lowest order for $t$-channel,
{\em i.e.} $e^- \mu^- \rightarrow e^- \mu^-$ process also diverges 
like in vacuum but in the presence of strong magnetic field, the logarithmic 
divergence in vacuum disappears and the infrared divergence remains.
As far as finite terms are concerned, the crosssection decreases 
faster like the $s$-channel. However, there is a negative 
term, which is independent of the center-of-mass energy and 
decreases with
the magnetic field. The above observation is also found 
in Bhabha and M\o ller scattering in the strong magnetic field.
The divergence in Bhabha scattering in vacuum arises at the lower 
limit of the incident angle whereas the same in the presence of
strong magnetic field arises due to the lower limit of transverse 
momentum. However the above divergences in the presence and in the absence
of strong magnetic field inter-related to each other
and we have derived the relation among these two      
divergences geometrically so that we can 
compare the crosssections in both cases at the same footing.

However, the above mentioned processes have also been studied extensively 
up to higher-order in vacuum ({\em i.e.} in the absence of magnetic 
field), as a result the divergences appeared have been controlled
by the higher-order corrections. For an example,
the collinear divergence in the Bhabha scattering 
had been cured by adding the corrections 
to the tree-level vacuum result by the radiative corrections 
in Ref.~\cite[and references therein]{G_montagna}, the $\mathcal{O} 
(\alpha^3)$ corrections~\cite{M_consoli}, the $\mathcal{O}(\alpha^4)$ 
corrections with the full mass dependence~\cite{bonciani_ferroglia}
etc., where  the contributions from the higher-order diagrams have been 
calculated and the IR and UV divergences were regularized by the dimensional 
regularization scheme.

\section {Acknowledgement}
We thank to Mr. Mujeeb Hasan, Mr. Bhaswar Chatterjee, Ms. Shubhalaxmi Rath and Mr.
Jitendra Pal for their help from time to time.


\appendix
\section{Matrix Element for electron-muon scattering: $s$-channel process}
This appendix contains detailed calculation of the squared matrix 
element for the process $e^-e^+\rightarrow \mu^-\mu^+$. 
We start with the matrix element for the Feynman diagram in Figure 
\ref{s_e_mu}: 
\begin{eqnarray}
\mathfrak{M}_{\rm s}\ems&=&\frac{-e^2}{q^2}\big[\overline{V}(y_B,k_{\omit{y}})\gamma^\mu U(y_A,p_{\omit{y}})\big]\big[\overline{U}(Y_D,K_{\omit{y}})\gamma_\mu V(Y_C,P_{\omit{y}})\big].\nonumber
\end{eqnarray}
In order to find the crosssection we have to take the 
square of the modulus of $\mathfrak{M}$ and then sum over the spin 
states
by factorizing into tensors at the electron (e) and muon (muon) 
vertex
\begin{equation}
\overline{\left|\mathfrak{M}_{\rm s}\right|^2}\ems=\frac{e^4}{q^4}L_{e}^{\mu\nu}L^{muon}_{\mu\nu},\label{A-matrix-element-square}
\end{equation}
where $L_{e}^{\mu\nu}$ and $L^{muon}_{\mu\nu}$
are given by 
\begin{eqnarray}
L_{e}^{\mu\nu}&=&\sum_{e~spins}\big[\overline{V}(y_B,k_{\omit{y}})\gamma^\mu U(y_A,p_{\omit{y}})\big]\big[\overline{V}(y_B,k_{\omit{y}})\gamma^\nu U(y_A,p_{\omit{y}})\big]^*,\nonumber\\
L^{muon}_{\mu\nu}&=&\sum_{\mu~ spins}\big[\overline{U}(Y_C,P_{\omit{y}})\gamma_\mu V(Y_D,K_{\omit{y}})\big]\big[\overline{U}(Y_C,P_{\omit{y}})\gamma_\nu V(Y_D,K_{\omit{y}})\big]^*.\nonumber
\end{eqnarray}
To calculate the complex conjugate of 
$\overline{V}(y_B,k)\gamma^\nu U(y_A,p)$, we start with the fact that  it 
is a $1\times1$ matrix therefore its complex conjugate is 
equal to its hermitian conjugate,
\begin{eqnarray}
\big[\overline{V}(y_B,k_{\omit{y}})\gamma^\nu U(y_A,p_{\omit{y}})\big]^*&=&\big[\overline{V}(y_B,k_{\omit{y}})\gamma^\nu U(y_A,p_{\omit{y}})\big]^\dag\nonumber\\
&=&\big[V^{\dag}(y_B,k_{\omit{y}})\gamma^0\gamma^\nu U(y_A,p_{\omit{y}})\big]^\dag\nonumber\\
&=&\big[U^{\dag}(y_A,p_{\omit{y}}){\gamma^\nu}^{\dag}{\gamma^0}^{\dag} V(y_B,k_{\omit{y}})\big]\nonumber\\
&=&\big[U^{\dag}(y_A,p_{\omit{y}})\gamma^0\gamma^\nu\gamma^0\gamma^0V(y_B,k_{\omit{y}})\big]\nonumber\\
&=&\big[\overline{U}(y_A,p_{\omit{y}})\gamma^\nu V(y_B,k_{\omit{y}})\big],\nonumber
\end{eqnarray}
where we have used some properties of gamma matrices like ${\gamma^\mu}^\dag=\gamma^0\gamma^\mu\gamma^0$ and $\gamma^0\gamma^0=I$. Applying these simplification the $L_e^{\mu\nu}$ becomes
\begin{eqnarray}
L_{e}^{\mu\nu}&=&\sum_{s,s'}\big[\overline{V}(y_B,k_{\omit{y}})\gamma^\mu U(y_A,p_{\omit{y}})\overline{U}(y_A,p_{\omit{y}})\gamma^\nu V(y_B,k_{\omit{y}})\big],\nonumber
\end{eqnarray}
where $s$ and $s'$ denote the spin states of electron and positron respectively.

To simplify $L^{\mu\nu}_e$, we begin with explicitly writing the above equation in terms of 
individual matrix elements, which tells us that the above equation is a 
$1\times 1$ matrix. Therefore we can calculate $L^{\mu\nu}_e$ by 
taking the trace of above equation. We then take the trace above 
equation and use the cyclic property of trace to form the 
completeness condition. Thus the above equation becomes 
\begin{eqnarray}
L_{e}^{\mu\nu}&=&Tr\Big[\sum_{s'} V^{s'}(y_B,k_{\omit{y}})\overline{V}^{s'}(y_B,k_{\omit{y}})\gamma^\mu \sum_{s}U^{s}(y_A,p_{\omit{y}})\overline{U}^{s}(y_A,p_{\omit{y}})\gamma^\nu\Big]\nonumber\\
&=&Tr\big[P_V(y_B,k_{\omit{y}})\gamma^\mu P_U(y_A,p_{\omit{y}})\gamma^\nu\big]\nonumber,
\end{eqnarray}
where $P_V$ and $P_U$ are the spin sums of the positron and the 
electron, 
respectively \cite{Furry,Bhattacharya:2007vz}. In strong 
magnetic field these spin sums are given by 
\begin{eqnarray}
P_V(y_B,n=0,\vec k_{\omit{y}})&=&\sum_{s'} V^{s'}(y_B,k_{\omit{y}})\overline{V}^{s'}(y_B,k_{\omit{y}})=\frac{1}{2}I_0^2(\xi_B)\big[-m(1-\Sigma_z)
+\slashed{k}_\parallel-\gamma_5\widetilde{\slashed{k}}_\parallel\big],
\nonumber\\
P_U(y_A,n=0,\vec p_{\omit{y}})&=&\sum_{s}U^{s}(y_A,p_{\omit{y}})\overline{U}^{s}(y_A,p_{\omit{y}})=\frac{1}{2}I_0^2(\xi_A)\big[m(1-\Sigma_z)
+\slashed{p}_\parallel-\gamma_5\widetilde{\slashed{p}}_\parallel\big],
\end{eqnarray}
where $\slashed{p}_\parallel=p^0\gamma^0-p^3\gamma^3$,  
$\widetilde{\slashed{p}}_\parallel=p^3\gamma^0-p^0\gamma^3$ and $m$ 
is the mass of electron with $I_0^2(\xi)=1$.

To further simplify $L_{e}^{\mu\nu}$ in a convenient way, let us denote $A=(1-\Sigma_z)$,  $K=\slashed{k}_\parallel-\gamma_5\widetilde{\slashed{k}}_\parallel$ and $P=\slashed{p}_\parallel-\gamma_5\widetilde{\slashed{p}}_\parallel$, and in this way $L_{e}^{\mu\nu}$ can be rewritten as
\begin{eqnarray*}
L_{e}^{\mu\nu}&=&\frac{1}{4}Tr\big[(-mA+K)\gamma^\mu(mA+P)\gamma^\nu\big]\\
&=&\frac{1}{4}Tr\bigg[-m^2(A\gamma^\mu A \gamma^\nu)+m(K\gamma^\mu A \gamma^\nu - A\gamma^\mu P \gamma^\nu)+(K\gamma^\mu P \gamma^\nu)\bigg].
\end{eqnarray*}
The coefficient of $m$ in the above equation contains odd number of 
gamma matrices, hence their trace vanishes and the above equation 
becomes
\begin{eqnarray}
	L_{e}^{\mu\nu}&=&\frac{1}{4}Tr\bigg[-m^2(A\gamma^\mu A \gamma^\nu)+(K\gamma^\mu P \gamma^\nu)\bigg].\label{L-e-mass}
\end{eqnarray}
Above equation contains two terms. We first simplify the first term 
$A\gamma^\mu A \gamma^\nu$ and calculate its trace. Using the 
value of $A=1-\Sigma_z$, we simplify the first term in the above 
equation as
\begin{eqnarray*}
A\gamma^\mu A \gamma^\nu&=&(1-\Sigma_z)\gamma^\mu (1-\Sigma_z) \gamma^\nu\\
&=&(1-i\gamma^1\gamma^2)\gamma^\mu (1-i\gamma^1\gamma^2) \gamma^\nu~~~~~\mbox{$(\Sigma_z=i\gamma^1\gamma^2)$}\\
&=&\gamma^\mu\gamma^\nu -i\gamma^\mu \gamma^1 \gamma^2 \gamma^\nu -i\gamma^1 \gamma^2 \gamma^\mu \gamma^\nu -\gamma^1 \gamma^2 \gamma^\mu \gamma^1 \gamma^2 \gamma^\nu,
\end{eqnarray*}
where the fourth term 
$(\gamma^1\gamma^2\gamma^\mu\gamma^1\gamma^2\gamma^\nu)$ in the 
above equation can be further simplified, using the property 
$\gamma^\mu\gamma^\nu+\gamma^\nu\gamma^\mu=2g^{\mu\nu}$, which is
\begin{equation*}
\gamma^1 \gamma^2 \gamma^\mu \gamma^1 \gamma^2 \gamma^\nu=2\gamma^{\mu_\perp}\gamma^{\nu}-\gamma^\mu\gamma^\nu,
\end{equation*}
where $\gamma^{\mu_\perp}=(0,\gamma^1,\gamma^1,0)$.

Thus $A\gamma^\mu A\gamma^\nu$ can be rewritten as
\begin{eqnarray*}
	A\gamma^\mu A \gamma^\nu&=&2\gamma^\mu\gamma^\nu -i\gamma^\mu \gamma^1 \gamma^2 \gamma^\nu -i\gamma^1 \gamma^2 \gamma^\mu \gamma^\nu -2\gamma^{\mu_\perp}\gamma^{\nu}.
\end{eqnarray*}
Using the trace properties {\em like} $Tr(\gamma^\mu \gamma^\nu 
\gamma^\lambda \gamma^\delta)=4\big[g^{\mu\nu}g^{\lambda\delta}-g^{\mu\lambda}g^{\nu\delta}+g^{\mu\delta}g^{\nu\lambda}\big]$ and 
$Tr(\gamma^\mu\gamma^\nu)=4g^{\mu\nu}$, the trace of $A\gamma^\mu A\gamma^\nu$ becomes
\begin{eqnarray}
Tr[A\gamma^\mu A \gamma^\nu]&=&8(g^{\mu\nu}-g^{\mu_{\perp}\nu}),
\end{eqnarray}
where $g^{\mu_{\perp}\nu}$ is 
$g^{\mu_{\perp}\nu_{\perp}}=(0,-1,-1,0)$. Therefore the above trace 
can be written as
\begin{eqnarray}
Tr[A\gamma^\mu A \gamma^\nu]&=&8(g^{\mu\nu}-g^{\mu_{\perp}\nu_{\perp}}).\label{L-e-mass-1}
\end{eqnarray}
The second term in~\eqref{L-e-mass} can be simplified as 
\begin{eqnarray}
K\gamma^\mu P\gamma^\nu&=&(\slashed{k}_\parallel-\gamma_5\widetilde{\slashed{k}}_\parallel)\gamma^\mu(\slashed{p}_\parallel-\gamma_5\widetilde{\slashed{p}}_\parallel)\gamma^\nu\nonumber\\
&=&(\slashed{k}_\parallel\gamma^\mu-\gamma_5\widetilde{\slashed{k}}_\parallel\gamma^\mu)(\slashed{p}_\parallel\gamma^\nu-\gamma_5\widetilde{\slashed{p}}_\parallel\gamma^\nu)\nonumber\\
&=&\slashed{k}_\parallel\gamma^\mu\slashed{p}_\parallel\gamma^\nu-\gamma_5\slashed{k}_\parallel\gamma^\mu\widetilde{\slashed{p}_\parallel}\gamma^\nu-\gamma_5\widetilde{\slashed{k}_\parallel}\gamma^\mu\slashed{p}_\parallel\gamma^\nu+\gamma_5\widetilde{\slashed{k}_\parallel}\gamma^\mu\gamma_5\widetilde{\slashed{p}_\parallel}\gamma^\nu \nonumber\\
&=&\slashed{k}_\parallel\gamma^\mu\slashed{p}_\parallel\gamma^\nu-\gamma_5\slashed{k}_\parallel\gamma^\mu\widetilde{\slashed{p}_\parallel}\gamma^\nu-\gamma_5\widetilde{\slashed{k}_\parallel}\gamma^\mu\slashed{p}_\parallel\gamma^\nu+\widetilde{\slashed{k}_\parallel}\gamma^\mu\widetilde{\slashed{p}_\parallel}\gamma^\nu . \label{A-electron-trace}
\end{eqnarray}
The above equation have four terms. The trace of the first and the last terms is given by the equations
\begin{eqnarray}
 Tr(\slashed{k}_\parallel\gamma^\mu\slashed{p}_\parallel\gamma^\nu)&=&4[p^\mu_\parallel k^\nu_\parallel+p^\nu_\parallel k^\mu_\parallel-(p_\parallel\cdot k_\parallel) g^{\mu\nu}],\\
 Tr(\widetilde{\slashed{k}}_\parallel\gamma^\mu\widetilde{\slashed{p}}_\parallel\gamma^\nu)&=&4[\widetilde p^\mu_\parallel \widetilde k^\nu_\parallel+\widetilde p^\nu_\parallel \widetilde k^\mu_\parallel-(\widetilde p_\parallel\cdot \widetilde k_\parallel) g^{\mu\nu}].
 \end{eqnarray}
For the trace of second and third term in~\eqref{A-electron-trace}, which are the standard traces, but to compare these with each other, we first expand the terms and 
then calculate the traces. The second term
in~\eqref{A-electron-trace} can be simplified as  
	\begin{eqnarray}
	\gamma_5\slashed{k}_\parallel\gamma^\mu\widetilde{\slashed{p}}_\parallel\gamma^\nu&=&\gamma_5(k^0\gamma^0-k^3\gamma^3)\gamma^\mu(p^0\gamma^3-p^3\gamma^0)\gamma^\nu\nonumber\\
	&=&\gamma_5\left[(k^0\gamma^0\gamma^\mu-k^3\gamma^3\gamma^\mu)(p^0\gamma^3\gamma^\nu-p^3\gamma^0\gamma^\nu)\right]\nonumber\\
	&=&\gamma_5[k^0p^0(\gamma^0\gamma^\mu\gamma^3\gamma^\nu)-k^0p^3(\gamma^0\gamma^\mu\gamma^0\gamma^\nu)-k^3p^0(\gamma^3\gamma^\mu\gamma^3\gamma^\nu)\nonumber\\
	&&\hspace{2mm}+k^3p^3(\gamma^3\gamma^\mu\gamma^0\gamma^\nu)]\nonumber\\
	&=&[k^0p^0(\gamma_5\gamma^0\gamma^\mu\gamma^3\gamma^\nu)-k^0p^3(\gamma_5\gamma^0\gamma^\mu\gamma^0\gamma^\nu)-k^3p^0(\gamma_5\gamma^3\gamma^\mu\gamma^3\gamma^\nu)\nonumber\\
	&&\hspace{2mm}+k^3p^3(\gamma_5\gamma^3\gamma^\mu\gamma^0\gamma^\nu)].\label{A-e-mu-trace}
	\end{eqnarray}
	Now we calculate the trace of above equation, which is
	\begin{eqnarray}
	Tr(\gamma_5\slashed{k}_\parallel\gamma^\mu\widetilde{\slashed{p}}_\parallel\gamma^\nu)&=&4i[k^0p^0\epsilon^{0\mu3\nu}-k^0p^3\epsilon^{0\mu0\nu}-k^3p^0\epsilon^{3\mu3\nu}+k^3p^3\epsilon^{3\mu0\nu}]\nonumber\\
	&=&4i[k^0p^0\epsilon^{0\mu3\nu}+k^3p^3\epsilon^{3\mu0\nu}]\nonumber\\
	&=&4i(k^0p^0-k^3p^3)\epsilon^{0\mu3\nu},\nonumber
	\end{eqnarray}
	where we use the trace property
	\begin{eqnarray}
	Tr(\gamma_5\gamma^\mu\gamma^\nu\gamma^\sigma\gamma^\lambda)&=&4i\epsilon^{\mu\nu\sigma\lambda},\nonumber\\
	\mbox{where}\hspace{2mm}\epsilon^{\mu\nu\sigma\lambda}&=&
	\begin{cases}
	1, & \text{for $\mu,\nu,\sigma,\lambda$ an even permutation of 0,1,2,3},\nonumber\\
	-1, & \text{for $\mu,\nu,\sigma,\lambda$ an odd permutation of 0,1,2,3},\nonumber\\
	0, & \text{if two indices are same}.
	\end{cases}
	\end{eqnarray}
	Similar to the second term, the third term in~\eqref{A-electron-trace} can be 
	simplified as
	\begin{eqnarray}
	\gamma_5\widetilde{\slashed{k}}_\parallel\gamma^\mu\slashed{p}_\parallel\gamma^\nu
	&=&\Big[k^0p^0(\gamma_5\gamma^3\gamma^\mu\gamma^0\gamma^\nu)-k^0p^3(\gamma_5\gamma^3\gamma^\mu\gamma^3\gamma^\nu)-k^3p^0(\gamma_5\gamma^0\gamma^\mu\gamma^0\gamma^\nu)\nonumber\\
	&&\hspace{2mm}+k^3p^3(\gamma_5\gamma^0\gamma^\mu\gamma^3\gamma^\nu)\Big].\nonumber
	\end{eqnarray}
	Thus the trace of above equation becomes
	\begin{eqnarray}
	Tr(\gamma_5\widetilde{\slashed{k}}_\parallel\gamma^\mu\slashed{p}_\parallel\gamma^\nu)&=&4i[k^0p^0\epsilon^{3\mu0\nu}-k^0p^3\epsilon^{3\mu3\nu}-k^3p^0\epsilon^{0\mu0\nu}+k^3p^3\epsilon^{0\mu3\nu}]\nonumber\\
	&=&4i[k^0p^0\epsilon^{3\mu0\nu}+k^3p^3\epsilon^{0\mu3\nu}]\nonumber\\
	&=&-4i(k^0p^0-k^3p^3)\epsilon^{0\mu3\nu}.\nonumber
	\end{eqnarray}
	
We can see that the trace of second and third term 
in~\eqref{A-electron-trace} cancel each other. Now we substitute 
the trace value of first and fourth term in 
Eq-\eqref{A-electron-trace}, which gives the trace 
of~\eqref{A-electron-trace} as
\begin{eqnarray}
Tr(K\gamma^\mu P\gamma^\nu)&=&4\Big[p^\mu_\parallel k^\nu_\parallel+p^\nu_\parallel k^\mu_\parallel-(p_\parallel\cdot k_\parallel) g^{\mu\nu}
+\widetilde p^\mu_\parallel \widetilde k^\nu_\parallel+\widetilde p^\nu_\parallel \widetilde k^\mu_\parallel-
(\widetilde p_\parallel\cdot \widetilde k_\parallel) g^{\mu\nu}\Big].\label{L-e-mass-2}
\end{eqnarray}
To get the final form of $L_e^{\mu\nu}$, we substitute the value from \eqref{L-e-mass-1} and 
\eqref{L-e-mass-2} in \eqref{L-e-mass}. Thus $L_e^{\mu\nu}$ becomes,
\begin{equation}\label{A-L-electron}
L_{e}^{\mu\nu}=\Big[p^\mu_\parallel k^\nu_\parallel+p^\nu_\parallel k^\mu_\parallel-(p_\parallel\cdot k_\parallel) g^{\mu\nu}
+\widetilde p^\mu_\parallel \widetilde k^\nu_\parallel+\widetilde p^\nu_\parallel \widetilde k^\mu_\parallel-
(\widetilde p_\parallel\cdot \widetilde k_\parallel) g^{\mu\nu}\Big]-2m^2(g^{\mu\nu}-g^{\mu_{\perp}\nu_\perp}).
\end{equation}
In a similar way the muonic vertex part can also be calculated as 
\begin{equation}\label{A-L-muon}
L^{muon}_{\mu\nu}=\Big[K_{\parallel \mu} P_{\parallel \nu}+K_{\parallel \nu} P_{\parallel \mu}-(K_\parallel\cdot P_\parallel) g_{\mu\nu}+
\widetilde K_{\parallel \mu} \widetilde P_{\parallel \nu}+\widetilde K_{\parallel \nu} \widetilde P_{\parallel \mu}-
(\widetilde K_\parallel\cdot \widetilde P_\parallel) g_{\mu\nu}\Big]-2M^2(g_{\mu\nu}-g_{\mu_{\perp}\nu_\perp}).
\end{equation}
Let us denote $L_{e}^{\mu\nu}=(T^{\mu\nu}+\widetilde{T}^{\mu\nu})$ 
and $L_{\mu\nu}^{muon}=(R_{\mu\nu}
+\widetilde{R}_{\mu\nu})$, where
\begin{eqnarray}
T^{\mu\nu}&=&p^\mu_\parallel k^\nu_\parallel+p^\nu_\parallel k^\mu_\parallel-(p_\parallel\cdot k_\parallel) g^{\mu\nu},\\
\widetilde{T}^{\mu\nu}&=&\widetilde p^\mu_\parallel \widetilde k^\nu_\parallel+\widetilde p^\nu_\parallel 
\widetilde k^\mu_\parallel-(\widetilde p_\parallel\cdot \widetilde k_\parallel) g^{\mu\nu},\\
R_{\mu\nu}&=&K_{\parallel \mu} P_{\parallel \nu}+K_{\parallel \nu} P_{\parallel \mu}-(K_\parallel\cdot P_\parallel) g_{\mu\nu},\\
\widetilde{R}_{\mu\nu}&=&\widetilde K_{\parallel \mu} \widetilde P_{\parallel \nu}+\widetilde K_{\parallel \nu} \widetilde P_{\parallel \mu}-
(\widetilde K_\parallel\cdot \widetilde P_\parallel) g_{\mu\nu}.
\end{eqnarray}
Therefore the spin summed squared matrix element~\eqref{A-matrix-element-square} becomes,
\begin{equation}\label{A-matrix-element-squre-2}
\overline{\left|\mathfrak{M}_{\rm s}\right|^2}\ems=\frac{e^4}{4q^4}(T^{\mu\nu}R_{\mu\nu}+T^{\mu\nu}\widetilde{R}_{\mu\nu}+\widetilde{T}^{\mu\nu}R_{\mu\nu}
+\widetilde{T}^{\mu\nu}\widetilde{R}_{\mu\nu})~,
\end{equation}
where we neglect the mass of 
electron as well as the mass of muon because of the reason that we 
are working in the 
extreme relativistic limit. Each term in the above equation can be 
calculated as
\begin{eqnarray}
T^{\mu\nu}R_{\mu\nu}&=&[p^\mu_\parallel k^\nu_\parallel+p^\nu_\parallel k^\mu_\parallel-(p_\parallel\cdot k_\parallel) g^{\mu\nu}][K_{\parallel \mu} P_{\parallel \nu}+K_{\parallel \nu} P_{\parallel \mu}-(K_\parallel\cdot P_\parallel) g_{\mu\nu}]\nonumber\\
&=&(p_\parallel\cdot K_\parallel)(k_\parallel\cdot P_\parallel)+(p_\parallel\cdot P_\parallel)(k_\parallel\cdot K_\parallel)-(p_\parallel\cdot k_\parallel)(K_\parallel\cdot P_\parallel)\nonumber\\
&&\hspace{1mm}+(p_\parallel\cdot P_\parallel)(k_\parallel\cdot K_\parallel)+(p_\parallel\cdot K_\parallel)(k_\parallel\cdot P_\parallel)-(p_\parallel\cdot k_\parallel)(K_\parallel\cdot P_\parallel)\nonumber\\
&&\hspace{1mm}-(p_\parallel\cdot k_\parallel)(K_\parallel\cdot P_\parallel)-(p_\parallel\cdot k_\parallel)(K_\parallel\cdot P_\parallel)+4(p_\parallel\cdot k_\parallel)(K_\parallel\cdot P_\parallel)\nonumber\\
&=&2\Big[(p_\parallel\cdot K_\parallel)(k_\parallel\cdot P_\parallel)+(p_\parallel\cdot P_\parallel)(k_\parallel\cdot K_\parallel)\Big].\nonumber
\end{eqnarray}
Similarly the other terms can also be calculated as
\begin{eqnarray}
\widetilde{T}^{\mu\nu}R_{\mu\nu}&=&2\Big[(\widetilde p_\parallel\cdot K_\parallel)(\widetilde k_\parallel\cdot P_\parallel)+(\widetilde 
p_\parallel\cdot P_\parallel)(\widetilde k_\parallel\cdot K_\parallel)\Big],\\
T^{\mu\nu}\widetilde{R}_{\mu\nu}&=&2\Big[(p_\parallel\cdot \widetilde K_\parallel)(k_\parallel\cdot \widetilde P_\parallel)+(p_\parallel\cdot 
\widetilde P_\parallel)(k_\parallel\cdot \widetilde K_\parallel)\Big],\\
\widetilde{T}^{\mu\nu}\widetilde{R}_{\mu\nu}&=&2\Big[(\widetilde p_\parallel\cdot \widetilde K_\parallel)(\widetilde k_\parallel\cdot 
\widetilde P_\parallel)+(\widetilde p_\parallel\cdot \widetilde P_\parallel)(\widetilde k_\parallel\cdot 
\widetilde K_\parallel)\Big].
\end{eqnarray}
Thus the spin summed matrix element becomes,
\begin{eqnarray}
\overline{\left|\mathfrak{M}_{\rm s}\right|^2}_{\rm B \neq 0}\ems&=&\frac{2e^4}{q^4}\Big[(p_\parallel\cdot K_\parallel)(k_\parallel\cdot P_\parallel)+(p_\parallel\cdot P_\parallel)(k_\parallel\cdot K_\parallel)+(p_\parallel\cdot \widetilde K_\parallel)(k_\parallel\cdot \widetilde P_\parallel)\nonumber\\
&&+(p_\parallel\cdot \widetilde P_\parallel)(k_\parallel\cdot \widetilde K_\parallel)+(\widetilde p_\parallel\cdot K_\parallel)(\widetilde k_\parallel\cdot P_\parallel)+(\widetilde p_\parallel\cdot P_\parallel)(\widetilde k_\parallel\cdot K_\parallel)\nonumber\\
&&+(\widetilde p_\parallel\cdot \widetilde K_\parallel)(\widetilde k_\parallel\cdot \widetilde P_\parallel)+(\widetilde p_\parallel\cdot \widetilde P_\parallel)(\widetilde k_\parallel\cdot \widetilde K_\parallel)\Big]. \label{A-matrix-element-square-3}
\end{eqnarray}
This can be rewritten using Mandelstam and Magnetic Mandelstam 
variables 
\begin{eqnarray}
\overline{\left|\mathfrak{M}_{\rm s}\right|^2}_{\rm B \neq 0}\ems&=&\frac{e^4}{2q^2}\Big[u^2+t^2+u_Ku_P+t_Pt_K+u_pu_k+t_pt_k+\widetilde{u}^2+\widetilde{t}^2\Big]\nonumber\\
&=&\frac{e^4}{s^2}\Big[u^2+t^2+u_pu_k+t_pt_k\Big].\label{A-matrix-element-square-4}
\end{eqnarray}

\section{Cross term in Bhabha Scattering}
We provide here the calculation of interference term of Bhabha 
scattering. We start from the matrix element for Bhabha scattering 
for $s$- and $t$-channel (figure~\ref{s-ch} and \ref{fd4})
\begin{eqnarray}
{\mathfrak{M}}_{\rm{s}}\bhabha 
&=& \frac{-e^2}{q_1^2}\big[\overline{V}(y_B,k_{\omit{y}})
\gamma^\mu U(y_A,p_{\omit{y}})\big]\big[\overline{U}(Y_C,P_{\omit{y}})
\gamma_\mu V(Y_D,K_{\omit{y}})\big], \\
\mathfrak{M}_{\rm{t}} (e^- e^+ \rightarrow e^- e^+)
&=& \frac{-e^2}{q_2^2}\big[\overline{U}(Y_C,P_{\omit{y}})
\gamma^\mu U(y_A,p_{\omit{y}})\big]\big[\overline{V}(y_B,k_{\omit{y}})
\gamma_\mu V(Y_D,K_{\omit{y}})\big].
\end{eqnarray}
Since the quantities in the square brackets are $1\times1$ matrices, 
we can rearrange them to form the completeness condition. 
Taking the sum 
over the spin states, the interference term, in the strong magnetic field limit, 
becomes 
\begin{eqnarray*}
	\overline{\mathfrak{M}_s\mathfrak{M}_t^*}&=&\frac{e^4}{q_1^2q_2^2}\sum_{all ~states}\big[\overline{V}(k_\parallel)\gamma^\mu U(p_\parallel)\big] \big[\overline{U}(p_\parallel)\gamma^\nu U(P_\parallel)\big] \big[\overline{U}(P_\parallel)\gamma_\mu V(K_\parallel)\big]\big[\overline{V}(K_\parallel)\gamma_\nu V(k_\parallel)\big] \\
	&=&\frac{e^4}{q_1^2q_2^2}Tr\Big[P_V(k_\parallel)\gamma^\mu P_U(p_\parallel)\gamma^\nu P_U(P_\parallel)\gamma_\mu P_V(K_\parallel)\gamma_\nu\Big]\\
	&=&\frac{e^4}{q_1^2q_2^2}Tr\big[(\slashed{k}_\parallel-\gamma_5\widetilde{\slashed{k}}_\parallel)\gamma^\mu (\slashed{p}_\parallel-\gamma_5\widetilde{\slashed{p}}_\parallel) \gamma^\nu (\slashed{P}_\parallel-\gamma_5\widetilde{\slashed{P}}_\parallel) \gamma_\mu (\slashed{K}_\parallel-\gamma_5\widetilde{\slashed{K}}_\parallel) \gamma_\nu\big]\\
	&=&\frac{e^4}{q_1^2q_2^2}Tr\big[(\slashed{k}_\parallel-\g_5 \tilde{\s{k}}_\parallel)\g^\mu (\s p_\parallel \g^\nu \s P_\parallel-\g_5 \s p_\parallel \g^\nu \tilde{\s P}_\parallel-\g_5 \tilde{\s p}_\parallel\g^\nu \s P_\parallel+\tilde{\s p}_\parallel\g^\nu \tilde{\s P}_\parallel) \g_\mu (\s{K}_\parallel-\g_5\tilde{\s{K}}_\parallel) \g_\nu\big],
\end{eqnarray*}
We further simplify the above equation by using the property 
$\gamma^\mu\slashed{a}\slashed{b}\slashed{c}\gamma_\mu=-2\slashed{c}\slashed{b}\slashed{a}$
\begin{eqnarray*}
	\overline{\mathfrak{M}_s\mathfrak{M}_t^*}&=&\frac{-2e^4}{q_1^2q_2^2}Tr\Big[(\slashed{k}_\parallel-\g_5 \tilde{\s{k}}_\parallel) (\s P_\parallel \g^\nu \s p_\parallel+\g_5 \tilde{\s P}_\parallel \g^\nu \s p_\parallel +\g_5 \s P_\parallel\g^\nu \tilde{\s p}_\parallel+\tilde{\s P}_\parallel\g^\nu \tilde{\s p}_\parallel)  (\s{K}_\parallel-\g_5\tilde{\s{K}}_\parallel) \g_\nu\Big]\\
	&=&\frac{-2e^4}{q_1^2q_2^2}Tr\Big[(\slashed{k}_\parallel-\g_5 \tilde{\s{k}}_\parallel) \Big(\s P_\parallel \g^\nu \s p_\parallel \s{K}_\parallel \g_\nu-\s P_\parallel \g^\nu \s p_\parallel \g_5\tilde{\s{K}}_\parallel \g_\nu+\g_5 \tilde{\s P}_\parallel \g^\nu \s p_\parallel \s{K}_\parallel \g_\nu \\
	&&\hspace{2cm}-\g_5 \tilde{\s P}_\parallel \g^\nu \s p_\parallel \g_5\tilde{\s{K}}_\parallel \g_\nu+\g_5 \s P_\parallel\g^\nu \tilde{\s p}_\parallel \s{K}_\parallel \g_\nu-\g_5 \s P_\parallel\g^\nu \tilde{\s p}_\parallel \g_5\tilde{\s{K}}_\parallel \g_\nu\\
	&&\hspace{2cm}+\tilde{\s P}_\parallel\g^\nu \tilde{\s p}_\parallel \s{K}_\parallel \g_\nu-\tilde{\s P}_\parallel\g^\nu \tilde{\s p}_\parallel \g_5\tilde{\s{K}}_\parallel \g_\nu\Big)\Big]\\
	&=&\frac{-2e^4}{q_1^2q_2^2}Tr\Big[(\slashed{k}_\parallel-\g_5 \tilde{\s{k}}_\parallel) \Big(\s P_\parallel \g^\nu \s p_\parallel \s{K}_\parallel \g_\nu+\g_5\s P_\parallel \g^\nu \s p_\parallel \tilde{\s{K}}_\parallel \g_\nu+\g_5 \tilde{\s P}_\parallel \g^\nu \s p_\parallel \s{K}_\parallel \g_\nu + \tilde{\s P}_\parallel \g^\nu \s p_\parallel \tilde{\s{K}}_\parallel \g_\nu\\
	&&\hspace{2cm}+\g_5 \s P_\parallel\g^\nu \tilde{\s p}_\parallel \s{K}_\parallel \g_\nu+ \s P_\parallel\g^\nu \tilde{\s p}_\parallel \tilde{\s{K}}_\parallel \g_\nu+\tilde{\s P}_\parallel\g^\nu \tilde{\s p}_\parallel \s{K}_\parallel \g_\nu+\g_5\tilde{\s P}_\parallel\g^\nu \tilde{\s p}_\parallel \tilde{\s{K}}_\parallel \g_\nu\Big)\Big].
\end{eqnarray*}
Using the property $\g^\mu\s{a}\s{b}\g_\mu=4a \cdot b$, the above 
equation can be further simplified
\begin{eqnarray*}
	\overline{\mathfrak{M}_s\mathfrak{M}_t^*}&=&\frac{-8e^4}{q_1^2q_2^2}Tr\Big[(\s{k}_\parallel-\g_5 \tilde{\s{k}}_\parallel) \Big\{\s P_\parallel (p_\parallel \cdot K_\parallel)+\g_5\s P_\parallel  (p_\parallel \cdot \tilde{K}_\parallel) +\g_5 \tilde{\s P}_\parallel  (p_\parallel \cdot K_\parallel)+ \tilde{\s P}_\parallel (p_\parallel \cdot \tilde{K}_\parallel) \\
	&&\hspace{2cm}+\g_5 \s P_\parallel (\tilde{p}_\parallel \cdot K_\parallel) + \s P_\parallel (\tilde{p}_\parallel\cdot \tilde{K}_\parallel) +\tilde{\s P}_\parallel (\tilde{p}_\parallel\cdot K_\parallel) +\g_5\tilde{\s P}_\parallel( \tilde{p}_\parallel\cdot \tilde{K}_\parallel) \Big\}\Big]\\
	&=&\frac{-8e^4}{q_1^2q_2^2}Tr\Big[\s{k}_\parallel\s P_\parallel (p_\parallel \cdot K_\parallel)-\g_5\s{k}_\parallel\s P_\parallel  (p_\parallel \cdot \tilde{K}_\parallel) -\g_5 \s{k}_\parallel\tilde{\s P}_\parallel  (p_\parallel \cdot K_\parallel)+ \s{k}_\parallel\tilde{\s P}_\parallel (p_\parallel \cdot \tilde{K}_\parallel) \\
	&&\hspace{1.5cm}-\g_5 \s{k}_\parallel\s P_\parallel (\tilde{p}_\parallel \cdot K_\parallel) + \s{k}_\parallel\s P_\parallel (\tilde{p}_\parallel\cdot \tilde{K}_\parallel) +\s{k}_\parallel\tilde{\s P}_\parallel (\tilde{p}_\parallel\cdot K_\parallel) -\g_5\s{k}_\parallel\tilde{\s P}_\parallel( \tilde{p}_\parallel\cdot \tilde{K}_\parallel)\\
	&&\hspace{1.5cm}-\g_5 \tilde{\s k}_\parallel\s P_\parallel (p_\parallel \cdot K_\parallel)+\tilde{\s k}_\parallel\s P_\parallel  (p_\parallel \cdot \tilde{K}_\parallel) + \tilde{\s k}_\parallel \tilde{\s P}_\parallel  (p_\parallel \cdot K_\parallel)- \g_5 \tilde{\s k}_\parallel\tilde{\s P}_\parallel (p_\parallel \cdot \tilde{K}_\parallel) \\
	&&\hspace{1.5cm}+ \tilde{\s k}_\parallel \s P_\parallel (\tilde{p}_\parallel \cdot K_\parallel) -\g_5 \tilde{\s k}_\parallel \s P_\parallel (\tilde{p}_\parallel\cdot \tilde{K}_\parallel) -\g_5 \tilde{\s k}_\parallel\tilde{\s P}_\parallel (\tilde{p}_\parallel\cdot K_\parallel) +\tilde{\s k}_\parallel\tilde{\s P}_\parallel( \tilde{p}_\parallel\cdot \tilde{K}_\parallel)\Big].
\end{eqnarray*}
To simplify the above equation, we use some trace property of gamma matrices {\em like} 
$Tr(\g_5 \s a \s b)=0$ and $Tr(\s a \s b)=4 a \cdot b$. Thus the 
above equation becomes
\begin{eqnarray*}
	\overline{\mathfrak{M}_s\mathfrak{M}_t^*} &=&\frac{-32e^4}{q_1^2q_2^2}\Big[(k_\parallel\cdot P_\parallel) (p_\parallel \cdot K_\parallel)+ (k_\parallel\cdot \tilde{P}_\parallel) (p_\parallel \cdot \tilde{K}_\parallel)+ (k_\parallel\cdot P_\parallel) (\tilde{p}_\parallel\cdot \tilde{K}_\parallel) +(k_\parallel\cdot \tilde{P}_\parallel) (\tilde{p}_\parallel\cdot K_\parallel) \\
	&&\hspace{1cm}+(\tilde{k}_\parallel\cdot P_\parallel)  (p_\parallel \cdot \tilde{K}_\parallel) + (\tilde{k}_\parallel \cdot \tilde{P}_\parallel)  (p_\parallel \cdot K_\parallel)+ (\tilde{k}_\parallel\cdot P_\parallel) (\tilde{p}_\parallel \cdot K_\parallel)+(\tilde{k}_\parallel\cdot \tilde{P}_\parallel)( \tilde{p}_\parallel\cdot \tilde{K}_\parallel)\Big].
\end{eqnarray*}
With the help of Mandelstam and magnetic Mandelstam variables, this 
can be further simplified
\begin{eqnarray*}
	\overline{\mathfrak{M}_s\mathfrak{M}_t^*}&=&\frac{-8e^4}{st}\Big[u^2+u_Ku_P+\widetilde{u}u+u_pu_P+u_ku_K+\widetilde{u}u+u_ku_p+\widetilde{u}^2\Big]\nonumber\\
	&=&\frac{-8e^4}{st}\Big[u_Ku_P+u_pu_P+u_ku_K+u_ku_p\Big]\nonumber\\
	&=&\frac{-8e^4}{st}\Big[u_Ku_P-u_Ku_P-u_Pu_K+u_Pu_K\Big]\nonumber\\
	&=&0.
\end{eqnarray*}

\section{Electron-Muon scattering in $s$ channel: $e^-e^+\rightarrow \mu^-\mu^+$}
This appendix contains the detailed calculation of crosssection for 
the process $\ems$. We start from the differential crosssection for 
the process $e^-e^+\rightarrow \mu^-\mu^+$ (Figure \ref{s_e_mu}) 
using 
the Eq-\eqref{liff} with the matrix 
element~\eqref{matrix-element-square-4}
\begin{eqnarray*}
	d\sigma^{\rm s}_{\rm B \neq 0}\ems&=&\frac{(2\pi)^4}{F} \overline{\left|\mathfrak{M}_{\rm s}\right|^2}\ems \delta^4(K_{\omit{y}}+P_{\omit{y}}-k_{\omit{y}}-p_{\omit{y}})\frac{d^3\vec{P}}{(2\pi)^32E_P}\frac{d^3\vec{K}}{(2\pi)^32E_K}\\
	&=&\frac{ \overline{\left|\mathfrak{M}_{\rm s}\right|^2}\ems}{(2\pi)^24F} \delta(E_P+E_K-E_p-E_k)\delta^3(\vec{K}_{\omit{y}}+\vec{P}_{\omit{y}}-\vec{k}_{\omit{y}}-\vec{p}_{\omit{y}})\\
	&&\hspace{10cm}\times\frac{d^3\vec{P}}{E_P}\frac{d^3\vec{K}}{E_K}.
\end{eqnarray*}
In center-of-mass frame, $\vec{p}+\vec{k}=0$, which also implies 
$\vec{p}_{\omit{y}}+\vec{k}_
{\omit{y}}=0$. Thus $d\sigma^{\rm s}_{\rm B \neq 0}\ems$ becomes
\begin{eqnarray*}
	d\sigma^{\rm s}_{\rm B \neq 0}\ems&=&\frac{ \overline{\left|\mathfrak{M}_{\rm s}\right|^2}\ems}{(2\pi)^24F} \delta(E_P+E_K-E_p-E_k)\delta^3(\vec{K}_{\omit{y}}+\vec{P}_{\omit{y}})\frac{d^3\vec{P}}{E_P}\frac{d^3\vec{K}}{E_K}\\
	&=&\frac{ \overline{\left|\mathfrak{M}_{\rm s}\right|^2}_{\vec{K}_{\omit{y}}=-\vec{P}_{\omit{y}}}\ems}{(2\pi)^24F} \delta(E_P+E_K-E_p-E_k)\frac{d^3\vec{P}}{E_KE_P}.
\end{eqnarray*}
In strong magnetic field $d^3\vec{P}$ split into $d^2P_
\perp$ and $dP_z$, where the integral over $d^2P_\perp$ in strong 
magnetic field limit becomes $\int_0^{|eB|}d^2P_\perp=\pi |eB|$.
This simplify the crosssection as
\begin{eqnarray}
\sigma^{\rm s}_{\rm B \neq 0} \ems&=&\int_{-\infty}^\infty \frac{\pi|eB| \overline{\left|\mathfrak{M}_{\rm s}\right|^2}_{K_z=-P_z}\ems}{(2\pi)^24F} \delta(E_P+E_K-E_p-E_k)\frac{dP_z}{E_KE_P}.\nonumber\\\label{C-d-sig} 
\end{eqnarray}
In strong magnetic field, $P_\perp \sim 0$, so 
the particles are restricted to move in the direction of magnetic field. They can either move 
in $+ve$ $z$ direction or in $-ve$ $ z$ direction.  Accordingly 
the four momentum dot product can be written as :-
\begin{equation}
p.k=
\begin{cases}
E_pE_k-|\vec{p}||\vec{k}|,& \mbox{if $\vec{p}$ and $ \vec{k}$ are in 
	the same direction},\\
E_pE_k+|\vec{p}||\vec{k}|,& \mbox{if $\vec{p}$ and $ \vec{k}$ are in 
	the opposite direction.} 
\end{cases}
\end{equation}
In extreme relativistic limit, first type of dot product is negligible 
compared
to the second type of dot product.
For 
$\theta=0$, $\theta$ is the scattering angle, the $t$ variables in eq~\eqref{matrix-element-square-4} 
are negligible as compared to all $u$ variables and for $\theta=180$
the $u$ variables are negligible as compared to all $t$ variables.

Thus for the one-dimensional scattering at high energy for 
$\theta=0$ degree
\begin{eqnarray}
\overline{\left|\mathfrak{M}_{\rm s}\right|^2}\ems&=&\frac{e^4}{4s^2}\Big[u^2+u_pu_k\Big],
\end{eqnarray} 
and for $\theta=180$ degree
\begin{eqnarray}
\overline{\left|\mathfrak{M}_{\rm s}\right|^2}\ems&=&\frac{e^4}{4s^2}\Big[t^2+t_pt_k\Big].
\end{eqnarray}
Let us denote $|\vec{p}|=|\vec{k}|=p_i$ and 
$|\vec{P}|=|\vec{K}|=p_f$ then $E_p=E_k=E_i$ 
and $E_K=E_P=E_f$. Using these energies and momenta, 
the value of $u$, $u_p$, $u_k$ can be calculated as       
\begin{eqnarray*}
	u&=&-2k\cdot P\\
	&=&-2[E_iE_f+p_ip_f].
\end{eqnarray*}
Therefore the square of $u$ becomes
\begin{eqnarray*}
	u^2&=&4[E_i^2E_f^2+p_i^2p_f^2+2E_iE_fp_ip_f].
\end{eqnarray*}
Similarly $u_p$, $u_k$ and their product become
\begin{eqnarray*}
	u_p&=&-2\widetilde{p}\cdot K=-2[\widetilde{p}^0K^0-\widetilde{p}^3K^3]\\
	&=&-2[p_zE_K+E_pK_z]\\
	&=&-2[p_iE_f+E_ip_f],\\
	u_k&=&-2\widetilde{k}\cdot P=-2[p_iE_f+E_ip_f],\\
	u_pu_k&=&4[p_i^2E_f^2+E_i^2p_f^2+2p_iE_fE_ip_f].
\end{eqnarray*}
The addition of $u^2$ and $u_pu_k$ can be simplified, using the approximation $p_f\sim E_f$, which holds good in the extreme relativistic limit. Thus $u^2+u_pu_k$ becomes
\begin{eqnarray*}
	u^2+u_pu_k&=&4\left[p_i^2E_f^2+E_i^2p_f^2+E_i^2E_f^2+p_i^2p_f^2+4p_iE_fE_ip_f \right]\\
	&=&4\Big[p_i^2E_f^2+E_i^2(E_f^2-M^2)+E_i^2E_f^2+p_i^2(E_f^2-M^2)\\
	&&+4p_i \sqrt{(E_f^2-M^2)}E_iE_f\Big]\\
	&=&4\left[p_i^2E_f^2+E_i^2E_f^2+E_i^2E_f^2+p_i^2E_f^2+4p_iE_fE_iE_f \right]\\
	&=&8E_f^2\left[p_i^2+E_i^2+4p_iE_i \right]\\
	&=&8E_f^2\left[E_i+p_i \right]^2.
\end{eqnarray*}
Thus for $\theta=0$ degree the square of the matrix element (which 
comes out to be same as for the $\theta=180$ degree) becomes 
\begin{equation}
\overline{\left|\mathfrak{M}_{\rm s}\right|^2}\ems=\frac{e^4}{4s^2}\Big[u^2+u_pu_k\Big]=\frac{2e^4}{s^2}E_f^2\left[E_i+p_i \right]^2,\label{C-ep_in_cm}
\end{equation}
and the flux factor for the collinear collision becomes
\begin{eqnarray*}
	F&=&|\vec{v}_p-\vec{v}_k|2E_p2E_k\\
	&=&\left[|\vec{v}_p|+|\vec{v}_k|\right]2E_p2E_k \qquad \mbox{(for collinear collision)}\\
	&=&4E_pE_k\left[\frac{|\vec{p}|}{E_p}+\frac{|\vec{k}|}{E_k}\right]\\
	&=&4\left[|\vec{p}|E_k+|\vec{k}|E_p\right]\\
	&=&4p_i\sqrt{s}. \hspace{4cm} (\mbox{$\sqrt{s}=E_p+E_k$})
\end{eqnarray*}
Threfore, using the expressions of $F$ and 
$\overline{\left|\mathfrak{M}\right|^2}$, eq-\eqref{C-d-sig} can be 
rewritten as
\begin{eqnarray}
	\sigma^{\rm s}_{\rm B \neq 0}\ems &=&\int_{-\infty}^{\infty}\frac{2\pi e^4|eB|E_f^2\left[E_i+p_i \right]^2}{s^2(2\pi)^216p_i\sqrt{s}} \delta(E_P+E_K-\sqrt{s})\frac{d\vec{P_z}}{E_f^2}\nonumber\\
	&=&2\int_{0}^{\infty}\frac{\pi e^4|eB|\left[E_i+p_i \right]^2}{s^2(2\pi)^2 8p_i\sqrt{s}} \delta(E_P+E_K-\sqrt{s})dp_f\label{C-d-sig1},
\end{eqnarray}
where the factor of $2$ comes due to the fact that the value of spin 
averaged squared matrix element is same for $P_z \in [0,\infty)$ 
and $P_z \in (-\infty,0]$ which corresponds to $\theta=0$ and 
$\theta=180$ degree respectively.

Let us denote $W=E_P+E_K$ which with the help of energy eigenvalue 
equation in the strong magnetic field limit becomes
\begin{equation*}
W=\sqrt{p_f^2+M^2}+\sqrt{p_f^2+M^2},
\end{equation*}
therefore the derivative $dW$ or $dp_f$ can be calculated as
\begin{eqnarray*}
	dW&=&\frac{p_fW}{E_PE_K} dp_f,\\
	\mbox{or,} \qquad dp_f&=&\frac{E_f^2}{p_fW}dW.
\end{eqnarray*}
To solve the Dirac Delta function we substitute the value of $dp_f$ 
in~\eqref{C-d-sig1}. Thus the crosssection becomes,
\begin{eqnarray*}
	\sigma^{\rm s}_{\rm B \neq 0}\ems &=&\int_{0}^{\infty}\frac{\pi e^4|eB|\left[E_i+p_i \right]^2}{s^2(2\pi)^2 4p_i\sqrt{s}} \delta(W-\sqrt{s})\frac{E_f^2}{p_fW}dW.
\end{eqnarray*}
With the help of the approximations: $p_f \approx E_f$ and $p_i \approx E_i$, the crosssection can be further simplified 
as
\begin{eqnarray*}
	\sigma^{\rm s}_{\rm B \neq 0} \ems&=&\int_{0}^{\infty}\frac{\pi e^4|eB|\left[2E_i \right]^2}{s^2(2\pi)^2 4E_i\sqrt{s}} \delta(W-\sqrt{s})\frac{E_f}{W}dW\\
	&=&\int_{0}^{\infty}\frac{\pi e^4|eB|\left[\sqrt{s} \right]^2}{s^2(2\pi)^2 2\sqrt{s}\sqrt{s}} \delta(W-\sqrt{s})\frac{W}{2W}dW\\
	&=&\int_{0}^{\infty}\frac{\pi e^4|eB|}{4s^2(2\pi)^2} \delta(W-\sqrt{s})dW\\
	&=&\frac{\pi e^4|eB|}{4s^2(2\pi)^2}.
\end{eqnarray*}
Using $e^2=4\pi \alpha$ the above result can be rewritten in terms of $\alpha$, which is 
\begin{eqnarray}
\sigma^{\rm s}_{\rm B \neq 0}\ems&=&\frac{\alpha^2\pi |eB|}{s^2}.
\end{eqnarray}

\section{Electron-Muon scattering in $t$ channel: $e^-\mu^-\rightarrow e^-\mu^-$}
This appendix provides the calculation of the crosssection for the 
$\emt$ process. We start from the fact that the 
matrix element for $t$ channel diagram has a pole at $t=0$, which 
can be easily justified by observing the equations below and 
the pole is arrived due to the lower limit of $P_\perp$, i.e. $P_\perp=0$ 
( $|\vec{P}|=\sqrt{P_\perp^2+P_z^2}$ and also 
$|\vec{p}|=|\vec{k}|=p_i$ and $|\vec{P}|=|\vec{K}|=p_f$).
\begin{eqnarray*}
	t^2=4|\vec{p}|^2|\vec{P}|^2(\cos \theta -1)^2=4|\vec{p}|^2|\vec{P}|^2\left(\frac{P_z}{|\vec{P}|}-1\right)^2=4p_i^2p_f^2\left(\frac{P_z}{p_f}-1\right)^2,\\
	u^2=4|\vec{p}|^2|\vec{K}|^2(\cos \theta +1)^2=4|\vec{p}|^2|\vec{K}|^2\left(\frac{P_z}{|\vec{P}|}+1\right)^2=4p_i^2p_f^2\left(\frac{P_z}{p_f}+1\right)^2.
\end{eqnarray*}
To deal with this problem, we apply a lower cut off, $\epsilon_B$ 
($\epsilon_B \to 0$) to 
$P_\perp$. Let us denote  $W=E_P+E_K$, $\sqrt{s}=E_p+E_k$, $E_p=E_k=E_i$ 
and 
$E_K=E_P=E_f$. Thus $d\sigma_{B \neq 0}^t$ can be written as
\begin{eqnarray*}
	d\sigma^{\rm{t}}_{\rm B \neq 0}\emt&=&\frac{ \overline{\left|\mathfrak{M}_t\right|^2}\emt}{(2\pi)^24F} \delta(W-\sqrt{s})\frac{P_\perp dP_\perp d\phi dP_z}{E_f^2},\\
	&=&\frac{e^4}{4t^2}\Big[s^2+u^2+s_ps_P+u_pu_P\Big]\frac{\delta(W-\sqrt{s})}{(2\pi)4F}\frac{P_\perp dP_\perp dP_z}{E_f^2}.
\end{eqnarray*}
The above equation, with the help of $t$ and $u$, can 
be rewritten as 
\begin{eqnarray*}
	d\sigma^{\rm{t}}_{\rm B \neq 0}&=&\hspace{-0.4cm}\bigintss_{-\infty}^\infty  dP_z\frac{e^4}{16p_i^2p_f^2\left(\frac{P_z}{p_f}-1\right)^2}\Big[s^2+4p_i^2p_f^2\left(\frac{P_z}{p_f}+1\right)^2+s_ps_P+u_pu_P\Big]\frac{\delta(W-\sqrt{s})}{(2\pi)4F}\frac{P_\perp dP_\perp}{E_f^2}.
\end{eqnarray*} 
The squared matrix element has different values for the different 
directions of $P_z$, so we split the integral into two parts, which 
gives 
\begin{eqnarray*}
	d\sigma^{\rm{t}}_{\rm B \neq 0}&=&\hspace{-0.4cm}\bigintss_{-\infty}^0  dP_z\frac{e^4}{16p_i^2p_f^2\left(\frac{P_z}{p_f}-1\right)^2}\Big[s^2+4p_i^2p_f^2\left(\frac{P_z}{p_f}+1\right)^2+s_ps_P+u_pu_P\Big]\frac{\delta(W-\sqrt{s})}{(2\pi)4F}\frac{P_\perp dP_\perp}{E_f^2}\\
	&+&\hspace{-0.4cm}\bigintss_{0}^\infty  dP_z\frac{e^4}{16p_i^2p_f^2\left(\frac{P_z}{p_f}-1\right)^2}\Big[s^2+4p_i^2p_f^2\left(\frac{P_z}{p_f}+1\right)^2+s_ps_P+u_pu_P\Big]\frac{\delta(W-\sqrt{s})}{(2\pi)4F}\frac{P_\perp dP_\perp}{E_f^2}.
\end{eqnarray*}
The first integral in the above equation, the $u$ variables are negligible 
compared to $s$ variables, thus $d\sigma^{\rm{t}}_{\rm B \neq 0}$ becomes,
\begin{eqnarray*}	
d\sigma^{\rm{t}}_{\rm B \neq 0}&=&\bigintss_{0}^\infty  dP_z\frac{e^4}{16p_i^2p_f^2\left(\frac{-P_z}{p_f}-1\right)^2}\Big[s^2+s_ps_P\Big]\frac{\delta(W-\sqrt{s})}{(2\pi)4F}\frac{P_\perp dP_\perp}{E_f^2}\\
	&+&\hspace{-0.2cm}\bigintss_{0}^\infty  dP_z\frac{e^4}{16p_i^2p_f^2\left(\frac{P_z}{p_f}-1\right)^2}\Big[s^2+4p_i^2p_f^2\left(\frac{P_z}{p_f}+1\right)^2+s_ps_P+u_pu_P\Big]\frac{\delta(W-\sqrt{s})}{(2\pi)4F}\frac{P_\perp dP_\perp}{E_f^2}.
\end{eqnarray*}
We can see that the second integral is the source of divergence 
and in a way we have separated the divergent term from the finite piece. 
After a little bit simplification, the above equation becomes 
\begin{eqnarray*}
	d\sigma^{\rm{t}}_{\rm B \neq 0}&=&\bigintss_{0}^\infty  dP_z\frac{e^4}{16p_i^2\left(-P_z-p_f\right)^2}\Big[s^2+s_ps_P\Big]\frac{\delta(W-\sqrt{s})}{(2\pi)4F}\frac{P_\perp dP_\perp}{E_f^2}\\
	&+&\bigintss_{0}^\infty  dP_z\frac{e^4}{16p_i^2\left(P_z-p_f\right)^2}\Big[s^2+4p_i^2\left(P_z+p_f\right)^2+s_ps_P+u_pu_P\Big]\frac{\delta(W-\sqrt{s})}{(2\pi)4F}\frac{P_\perp dP_\perp}{E_f^2}.
\end{eqnarray*}
As we discussed earlier that $P_\perp \sim 0$ in the strong magnetic field. Therefore the above differences in the momenta can be simplified by the
approximations: $P_z-p_f\simeq-\frac{P_\perp^2}{2P_z}$ and 
$P_z+p_f\simeq 
2P_z$, hence the above integral can be rewritten as
\begin{eqnarray*}
	d\sigma^{\rm{t}}_{\rm B \neq 0}&=&\bigintss_{0}^\infty  dP_z\frac{e^4}{64p_i^2P_z^2}\Big[s^2+s_ps_P\Big]\frac{\delta(W-\sqrt{s})}{(2\pi)4F}\frac{P_\perp dP_\perp}{E_f^2}\\
	&+&\bigintss_{0}^\infty  dP_z\frac{e^4P_z^2}{4p_i^2P_\perp^4}\Big[s^2+16p_i^2P_z^2+s_ps_P+u_pu_P\Big]\frac{\delta(W-\sqrt{s})}{(2\pi)4F}\frac{P_\perp dP_\perp}{E_f^2}.
\end{eqnarray*}
We have separated the integral in functions of $P_z$ and 
$P_\perp$. Next we integrate over $P_\perp$, by applying a 
lower cut off to the $P_\perp$ where it causes the divergence. Thus 
the above equation simplifies as
\begin{eqnarray*} 
	\sigma^{\rm{t}}_{\rm B \neq 0}&=&\bigintss_{0}^\infty  dP_z\frac{e^4}{64p_i^2P_z^2}\Big[s^2+s_ps_P\Big]\frac{\delta(W-\sqrt{s})}{(2\pi)4 F E_f^2}\bigintss_{0}^{\sqrt{|eB|}} P_\perp dP_\perp\\
	&+&\bigintss_{0}^\infty  dP_z\frac{e^4P_z^2}{4p_i^2}\Big[s^2+16p_i^2P_z^2+s_ps_P+u_pu_P\Big]\frac{\delta(W-\sqrt{s})}{(2\pi)4 F E_f^2} \lim_{\epsilon_B \to 0}\bigintss_{\epsilon_B}^{\sqrt{|eB|}}\frac{dP_\perp}{P_\perp^3},\\
	&=&\bigintss_{0}^\infty  dP_z\frac{e^4}{64p_i^2P_z^2}\Big[16E_i^4+16k_zE_iK_zE_f\Big]\frac{\delta(W-\sqrt{s})}{(2\pi)4 F E_f^2}\frac{|eB|}{2}\\
	&+&\bigintss_{0}^\infty  dP_z\frac{e^4P_z^2}{4p_i^2}\Big[16E_i^4+16p_i^2P_z^2+16k_zE_iK_zE_f+u_pu_P\Big]\frac{\delta(W-\sqrt{s})}{(2\pi)4 F E_f^2} 
\lim_{\epsilon_B \to 0}\left[\frac{1}{\epsilon^2_B}-\frac{1}{|eB|}\right].
\end{eqnarray*}
Since the perpendicular component of momentum is very small 
compared to the $z$ 
component of momentum, we approximate $P_z=K_z\simeq p_f$.
Therefore $\sigma^{\rm{t}}_{\rm B \neq 0}$ becomes
\begin{eqnarray*}
	\sigma^{\rm{t}}_{\rm B \neq 0}&=&\bigintss_{0}^\infty  dp_f\frac{e^4}{64p_i^2p_f^2}\Big[16E_i^4+16p_iE_ip_fE_f\Big]\frac{\delta(W-\sqrt{s})}{32\pi p_i\sqrt{s} E_f^2}\frac{|eB|}{2}\\
	&+&\bigintss_{0}^\infty  \left[dp_f\frac{e^4p_f^2}{4p_i^2}\Big\{16E_i^4+16p_i^2p_f^2+16p_iE_ip_fE_f+4(p_iE_f+E_ip_f)^2\Big\}\right.\\
	&&\left.\hspace{4cm}\times\frac{\delta(W-\sqrt{s})}{32\pi p_i\sqrt{s} E_f^2} \lim_{\epsilon_B \to 0}\left\{\frac{1}{\epsilon^2_B}-\frac{1}{|eB|}\right\}\right].
\end{eqnarray*}
To deal with the Dirac Delta function, we write $dp_f$ in terms of 
$dW$ i.e.  
$dp_f=\frac{E_f^2}{p_fW}dW$ and approximate $p_i\simeq E_i$ and 
$p_f\simeq E_f$. Thus the above integral becomes 
\begin{eqnarray*}
	\sigma^{\rm{t}}_{\rm B \neq 0}&=&\bigintss_{0}^\infty  dW\frac{e^4}{8 E_f W}\Big[E_i^2+E_f^2\Big]\frac{\delta(W-\sqrt{s})}{32\pi E_i\sqrt{s} E_f^2}|eB|\\
	&+&\bigintss_{0}^\infty  \frac{E_f}{W}dW e^4\Big[E_i^2+3E_f^2\Big]\frac{\delta(W-\sqrt{s})}{8\pi E_i\sqrt{s}} \lim_{\epsilon_B \to 0}\left[\frac{1}{\epsilon^2_B}-\frac{1}{|eB|}\right].
\end{eqnarray*}
Using $E_f=W/2$ and $E_i=\sqrt{s}/2$, the above integral can be 
further simplified as
\begin{eqnarray*}
	\sigma^{\rm{t}}_{\rm B \neq 0}&=&\bigintss_{0}^\infty  dW\frac{e^4}{16W^2}\Big[s+W^2\Big]\frac{\delta(W-\sqrt{s})}{4\pi s W^2}|eB|\\
	&+&\bigintss_{0}^\infty  dW e^4\Big[s+3W^2\Big]\frac{\delta(W-\sqrt{s})}{32\pi s} \lim_{\epsilon_B \to 0}\left[\frac{1}{\epsilon^2_B}-\frac{1}{|eB|}\right].
\end{eqnarray*}
To obtain the crosssection, we integrate over $dW$ by using the property of Dirac Delta 
function. Thus the above equation becomes
\begin{eqnarray*}
	\sigma^{\rm{t}}_{\rm B \neq 0}&=&\frac{e^4}{32\pi s^2}|eB|+\frac{e^4}{8\pi} \lim_{\epsilon_B \to 0}\left[\frac{1}{\epsilon^2_B}-\frac{1}{|eB|}\right]\\
	&=&\frac{\pi \alpha^2}{2s^2}|eB|-\frac{2\pi \alpha^2}{|eB|}+2\pi \alpha^2\lim_{\epsilon_B \to 0}\left[\frac{1}{\epsilon^2_B}\right].
\end{eqnarray*}

\noindent {\bf Relation between the divergence of vacuum and magnetic field} \\
Starting from the relation $P_\perp=P_z \tan \theta$, where 
$P_\perp$ is the momentum of the particle in the transverse direction, 
we can calculate
	\begin{eqnarray*}
		\frac{1}{P^2_\perp}=\frac{1}{P_z^2\tan^2\theta}=\frac{1}{P_z^2}\left[\frac{\cos^2\theta}{\sin^2\theta}\right].
	\end{eqnarray*}
	Let us denote $\cos\theta=x$, which simplify the above equation as
	\begin{eqnarray*}
		\frac{1}{P^2_\perp}&=&\frac{1}{P_z^2}\left[\frac{x^2}{1-x^2}\right]\\
		&=&\frac{x^2}{2P_z^2}\left[\frac{1}{1+x}+\frac{1}{1-x}\right]\\
		&=&\frac{1}{2P_z^2}\left[\frac{x^2}{1+x}+\frac{x^2}{1-x}\right]\\
		&=&\frac{1}{2P_z^2}\left[\frac{x^2}{1+x}+\frac{x^2-1+1}{1-x}\right]\\
		&=&\frac{1}{2P_z^2}\left[\frac{x^2}{1+x}-(1+x)+\frac{1}{1-x}\right].
	\end{eqnarray*}
	We set a lower cut off to $\theta$, $\theta_0 \to 0$, which gives 
	$x_0 \to 1$, for $x_0=\cos\theta_0$. Thus above equation becomes
	\begin{eqnarray*}
		\frac{1}{P^2_\perp}&=&\frac{1}{2P_z^2}\left[\frac{1}{1+1}-(1+1)+\lim_{x_0 \to 1}\frac{1}{1-x_0}\right]\\
		&=&\frac{1}{2P_z^2}\left[-\frac{3}{2}+\lim_{x_0 \to 1}\frac{1}{1-x_0}\right],
	\end{eqnarray*}
where we set $x_0=1$ in those terms which don't cause the divergence.
 
In the strong magnetic field $|\vec{P}|\approx P_z$ and in the 
extreme relativistic limit, $P_z \approx E_P$. With the help of 
these approximations, we can approximate $P_z^2\approx s/4$i, which
thus simplifies the above equation as 
	\begin{eqnarray*}
		\frac{1}{P^2_\perp}&=&\frac{2}{s}\left[-\frac{3}{2}+\lim_{x_0 \to 1}\frac{1}{1-x_0}\right]\\
		&=&\frac{2}{s}\left[-\frac{3}{2}+\lim_{\cos\theta_0 \to 1}\frac{1}{1-\cos\theta_0}\right],\\
		\frac{1}{\epsilon^2_B}&=&\frac{2}{s}\left[-\frac{3}{2}+\frac{1}{\epsilon_V}\right],\\
		&or&\\
		\frac{1}{\epsilon_V}&=&\frac{s}{2\epsilon_B^2}+\frac{3}{2},
	\end{eqnarray*}
where $\epsilon_B$ is the lower cut off on $P_\perp$ and 
$\epsilon_V=1-\cos \theta_0$.

\end{document}